\documentclass{iau-JDSS}
\usepackage{graphicx}

\title[Eta Carinae] 
{Eta Carinae in the Context of the Most Massive Stars}

\author[Gull \& Damineli]  
{Theodore R. Gull$^1$ \and Augusto Damineli$^2$}

\affiliation{$^1$Laboratory for Extraterrestial Planets and Stellar Astrophysics, Code 667, NASA/GSFC, Greenbelt, MD, 20771, USA, email: Theodore.R.Gull@nasa.gov\\$^2$IAGUSP, Universidade de Sao Paulo, 
Rua do Matao 1226, Sao Paulo, 05508-900, Brazil, email:
damineli@astro.iag.usp.br\\[\affilskip]
}

\pubyear{2009}
\volume{Volume 15}  
\pagerange{1--30}
\date{Sep 30, 2009 and in revised form ??}
\jname{Highlights of Astronomy, Volume 14}
\editors{Ian F Corbett, ed.}
\begin{document}
\maketitle

\firstsection
Eta Car, with its historical outbursts, visible ejecta and massive, variable winds, continues to challenge both observers and modelers. In just the past five years over 100 papers have been published on this fascinating object. We now know it to be a massive binary system with a 5.54-year period. In January 2009, Eta \ Car underwent one of its periodic low-states, associated with periastron passage of the two massive stars. This event was monitored by an intensive multi-wavelength campaign ranging from $\gamma$-rays to radio. A large amount of data was collected to test a number of evolving models including 3-D models of the massive interacting winds. August 2009 was an excellent time for observers and theorists to come together and review the accumulated studies, as have occurred in four meetings since 1998 devoted to Eta Car. Indeed, Eta \ Car behaved both predictably and unpredictably during this most recent periastron, spurring timely discussions.\\\indent
Coincidently, WR140  also passed through periastron in early 2009. It, too, is a intensively studied massive interacting binary. Comparison of its properties, as well as the properties of other massive stars,  with those of Eta Car is very instructive. These well-known examples of evolved massive binary systems provide many clues as to the fate of the most massive stars.\\\indent What are the effects of the interacting winds, of individual stellar rotation, and of the circumstellar material on what we see as hypernovae/supernovae? We hope to learn.\\
\indent Topics discussed in this 1.5 day Joint Discussion were:
\\ Eta \ Car: the 2009.0 event: Monitoring campaigns in X-rays, optical, radio, interferometry
\\WR140 and HD5980: similarities and differences to Eta \ Car
\\LBVs and Eta Carinae: What is the relationship?
\\Massive binary systems, wind interactions and 3-D modeling
\\Shapes of the Homunculus \& Little Homunculus: what do we learn about mass ejection?
\\Massive stars: the connection to supernovae, hypernovae and gamma ray bursters
\\Where do we go from here? (future directions)\\
\indent The Science Organizing Committee:\\Co-chairs: Augusto Damineli (Brazil) \& Theodore R. Gull (USA). Members:
D. John Hillier (USA),
Gloria Koenigsberger (Mexico), Georges Meynet (Switzerland),
Nidia I. Morrell (Chile),
Atsuo T. Okazaki (Japan),
Stanley P. Owocki (USA),
Andy M.T. Pollock (Spain),
Nathan Smith (USA),
Christiaan L. Sterken (Belgium),
Nicole St Louis (Canada),
Karel A. van der Hucht (Netherlands),
Roberto Viotti (Italy) and
Gerd Weigelt (Germany)\\
\\Website for talks and posters:\\
http://astrophysics.gsfc.nasa.gov/research/etacar/IAUJD.html
\section{Oral Presentations}
\subsection{Dedication to Prof. Sveneric Johansson  (Henrik Hartman)} 
Professor Sveneric Johansson is remembered for his important contributions to the know- ledge on atomic data, focusing on the iron group elements in general and singly ionized iron, Fe\,{\sc ii}, in particular. His work includes term analysis of several important ions, and measurements of atomic parameters for astrophysicaly important elements. His thorough knowledge of atomic structure also allowed major contributions to the analysis of complex astronomical spectra and atomic photo processes. 
Sveneric is greatly missed as an ingenious scientist, positive colleague and a great friend.\\\begin{figure}[t]
\includegraphics[width=5.2in,angle=0]{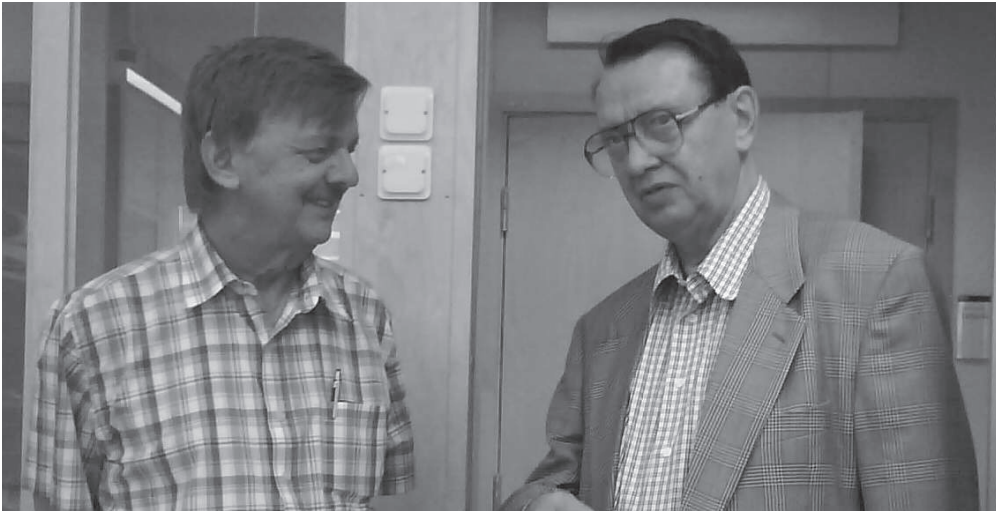}
  \caption{Professors Sveneric Johansson and Vladelin Letokhov discussing the stimulated emission properties of the ionized ejecta surrounding Eta \ Car. Both researchers passed away this past year. Their interest in atomic spectroscopy and enthusiasm was infectious to all.}
\end{figure}
\indent Sveneric received his PhD from Lund University in 1978 under the supervision of Professor Bengt Edl\'en, on the subject of term analysis of Fe\,{\sc ii} (the spectrum of Fe$^+$). This work continued to be his main research topic for more than 35 years. Sveneric led classical atomic spectroscopy into a new era of measurements with crucial astronomical applications. He spent a sabbatical year at NASA's Goddard Space Flight Center with Dave Leckrone during 1987-1988, starting up a collaboration for the upcoming Hubble Space Telescope (HST) mission and the $\chi$~Lupi pathfinder project. The high resolution spectrographs onboard HST challenged  existing laboratory atomic data bases. Sveneric foresaw the need of high-accuracy ultraviolet data and directed, together with Ulf Litz\'en, the Lund University spectroscopy laboratory to measure wavelengths, isotopic shifts and line structures needed to interpret astronomical observations. 
Spectroscopic investigations included iron, yttrium, mercury, boron, gold, ruthenium, nickel, thallium, platinum, and zirconium.\\
\indent The high cosmic abundance of iron makes  Fe\,{\sc ii} lines abundant in a variety of astronomical objects. For quantitative analyses the intrinsic strength of the spectral lines need to be known. In 2001 Sveneric founded the Atomic Astrophysics group at Lund University and organized the FERRUM project, an international collaboration on oscillator strengths for iron group elements. The aim of this project is to present a fully evaluated and consistent set of values, experimental and theoretical, that can be used for astronomical analyses.\\
\indent Throughout his career Sveneric also analyzed complex astronomical emission line spectra, and was especially interested in atomic photo processes. Together with Professor Vladilen Letokhov he identified and developed the idea of stimulated emission (LASER) in gas condensations close to the massive star Eta Carinae. From the strange behavior observed and ionization structure of the high ionization lines, they derived the concept of resonance-enhanced two-photon ionization (RETPI) of Ne and Ar atoms as an explanation for the production of these ions.\\\indent
In addition, it is with great sadness, that we learnt of Dr. Vladelen Letokhov's passing during 2009. He is greatly missed by colleagues and friends all over the world. 
During his productive career he published nearly 900 articles and 16 monographs. 
Sveneric's and Vladilen's work on photo processes culminated in their book '\textit{Astrophysical Lasers}' (Oxford Press, 2009) published earlier this year. 
\subsection{The historical background on Eta Car  (D. John Hillier)}
Eta Carinae, a spectacular object, is one of the most luminous stars in the galaxy,
and exhibits a wide range of interesting phenomena with implications for many areas of astrophysics. In this presentation we provide a brief summary of key discoveries and an introduction to some jargon associated with Eta~Car. 
\\\begin{figure}[ht]
\includegraphics[width=5.3in,angle=0]{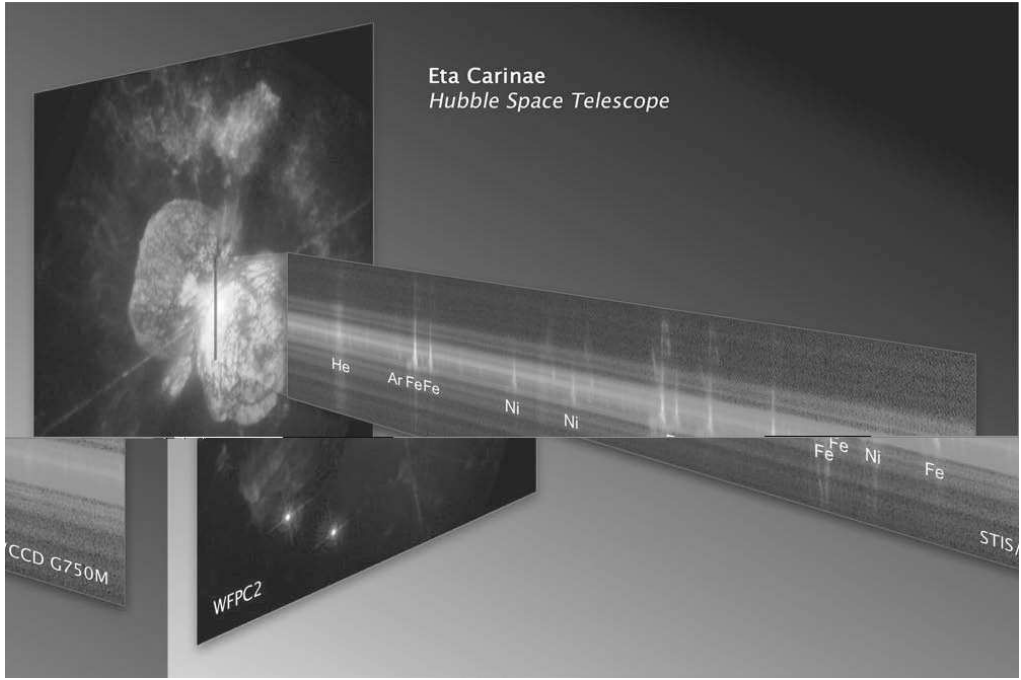}
  \caption{An example of what is so intriguing about Eta  Car: the extended wind and ejecta. A 0.1"-wide slit of the Hubble Space Telescope Imaging Spectrograph samples the extended structure surrounding Eta \ Car as imaged by Hubble Space Telescope.  Continuum and broad line emission at the center of the spectrum originate from the extended interacting winds. Narrow forbidden emission lines shifted with velocities up to 500 km s$^{-1}$ come from the interior of the Homunculus, thrown out in the 1840s. An estimated 10 to 20 $M_\odot$ was ejected during the Great Eruption as well as up to 0.5 $M_\odot$ in the lesser eruption of the 1890s. How did the ejecting star survive and what clues does this provide us on the late stages of massive stellar evolution? (Image courtesy of NASA and STScI)}
\end{figure}
Eta Carinae, a spectacular object, is one of the most luminous stars in the galaxy,
and exhibits a wide range of interesting phenomena with implications for many areas of astrophysics. In this presentation we provide a brief summary of key discoveries and an introduction to some jargon associated with Eta ~Carinae. \\
\indent In the 1840's Eta ~Carinae underwent a giant outburst and ejected a nebula which
we call the Homunculus. The event was so impressive that Eta ~Carinae was classified as a peculiar SN. With the onset of dust formation, it suffered a dramatic drop in brightness by $\sim 6$ magnitudes (e.g., van Genderen et~al. 1984, Space Sci. Rev., 39, 317). In the early 1890's Eta ~Carinae underwent a smaller outburst ejecting the Little Homunculus nebula (discovered with the HST; Ishibashi et~al. 2003, AJ, 125, 3222). \\
\indent The Homunculus is a bipolar nebula whose axis is tilted at about 41$^{\circ}$ to our line of sight. H$_2$ emission and dust is confined to a thin outer layer, while [Fe\,{\sc ii}] \& [Ni\,{\sc ii}] emission lines originate inside this shell
(Smith et~al. 2006, ApJ, 644, 1151). From infrared observations the mass of the Homunculus is inferred to exceed 10$M_{\odot}$ (Smith et~al, 2003, AJ, 125, 1458), and is possibly as large as 20$M_{\odot}$ (Smith et al. 2007, ApJ, 655, 911). In contrast, the mass of the Little Homunculus is $\sim 0.1M_ {\odot}$ (Smith 2005, MNRAS, 357, 1330). \\
\indent S-condensation ejecta (a condensation to the south of the Homunculus)  are N enhanced and CO depleted, consistent with the influence of CNO processing (Davidson et~al. 1982, ApJ, 254, L47). A similar abundance pattern is seen in the star (Hillier et~al. 2001, ApJ, 553, 837). As Eta ~Carinae is located in a region of massive star formation (Walborn et~al. 1977, ApJ, 211, 181), it is inferred that it is a young, but evolved, massive star.\\
\indent Speckle observations showed that Eta ~Carinae is composed of 4 `star-like' objects (Weigelt et al. 1986, A\&A, 163, L5). Subsequent HST observations revealed that the brightest of these is truly star-like, while the remaining 3 are  small nebula  (the Weigelt blobs) which emit the narrow permitted and forbidden lines that are prominent in ground-based spectra (Davidson et al. 1995, AJ, 109, 1784); they are prominent because the primary star suffers additional extinction ($\sim$\,5 magnitudes in 1997; Hillier et~al. 2001, ApJ, 553, 837). \\
\indent The discovery of a 5.5 year variability cycle (Damineli 1996, ApJ, 460, L49) led to the realization that Eta ~Carinae is a binary system (Damineli et al. 1997, New Astr., 2, 107). A wide range of phenomena, including infrared (Whitelock et~al. 2004, MNRAS, 352, 447), X-ray (Ishibashi et~al. 1999, ApJ, 524, 983; Corcoran 2005, AJ, 129, 2018), radio (Duncan et~al. 1999, ASP Conf. Ser. 179, 54), and line variability (Damineli et~al. 2008, MNRAS, 386, 2330) indicate that we are dealing with a binary system with a large orbital eccentricity ($\epsilon \sim 0.9$).\\
\indent The spectrum of the primary is similar to the P Cygni star HDE 316285. Modeling places a lower limit of 60$R_{\odot}$ on the radius of the central star, although with a re-interpretation of the He\,{\sc i} emission lines  a larger radius ($\sim 240R_{\odot}$) is now preferred. Because of the very dense wind we observe the wind --- not the ``normal'' photosphere of the star ($\dot M \sim 10^{-3}$ $M_{\odot}$/yr; Hillier et~al. 2001, ApJ, 553, 837).\\
\indent UV spectra reveal multiple systems of narrow absorption lines arising from neutral and singly ionized metals, and from H$_2$ (Gull et al. 2005, ApJ, 620, 442). The two dominant systems are associated with the Little Homunculus and the Homunculus, with other systems thought to be related to structures arising from the periodic interaction between the winds of the primary and secondary stars.\\
\indent HST observations show that the central star has brightened -- by over a factor of 3 since the first HST observations (Martin et al. 2004, AJ, 127, 2352). This is presumably due to a reduction in extinction, since spectra of the star, and the Weigelt blobs, have not shown dramatic changes.
Variability observations show that spectral changes occur throughout the 5.5 year cycle. This provides additional evidence for binarity; the variability most likely arises from illumination effects of the Weigelt blobs as the secondary star (believed to be responsible for ionizing the Weigelt blobs) moves in its orbit.
\\
\indent HST observations show that the broad He{\,\sc i} emission  lines most likely originate in the neighborhood of the wind-wind interface, and are not excited by the primary star. They exhibit complex radial velocity and profile variations which are broadly consistent with those expected in a binary system (Nielsen et al. 2007, ApJ, 660, 669). 	
\subsection{The 2009 monitoring campaign}
\subsubsection{The X-ray light curve (Michael F. Corcoran \& Kenji Hamaguchi)}
\begin{figure}[ht]
\begin{center}
 \includegraphics[width=2.2in]{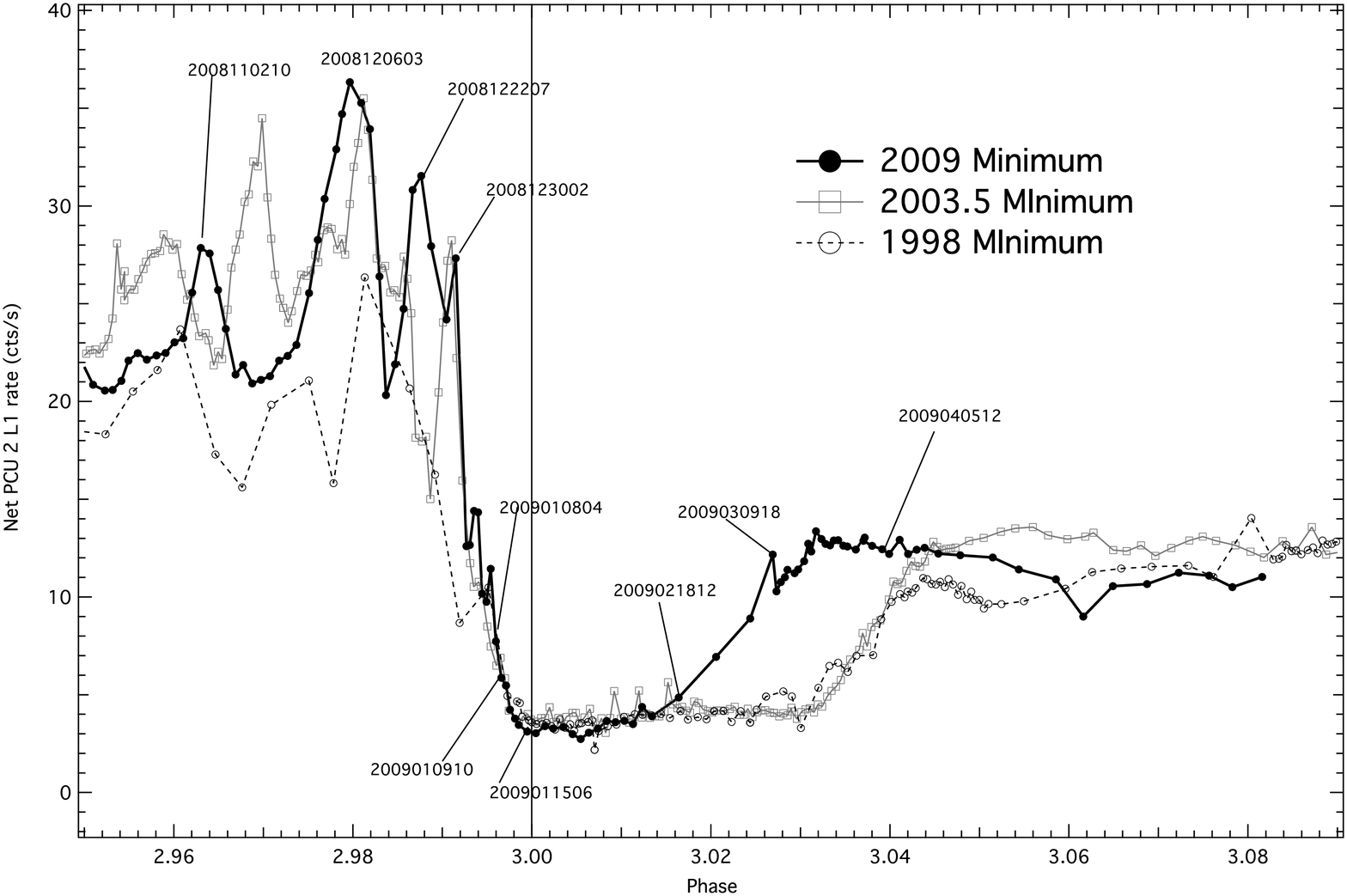} 
 \includegraphics[width=3.0in]{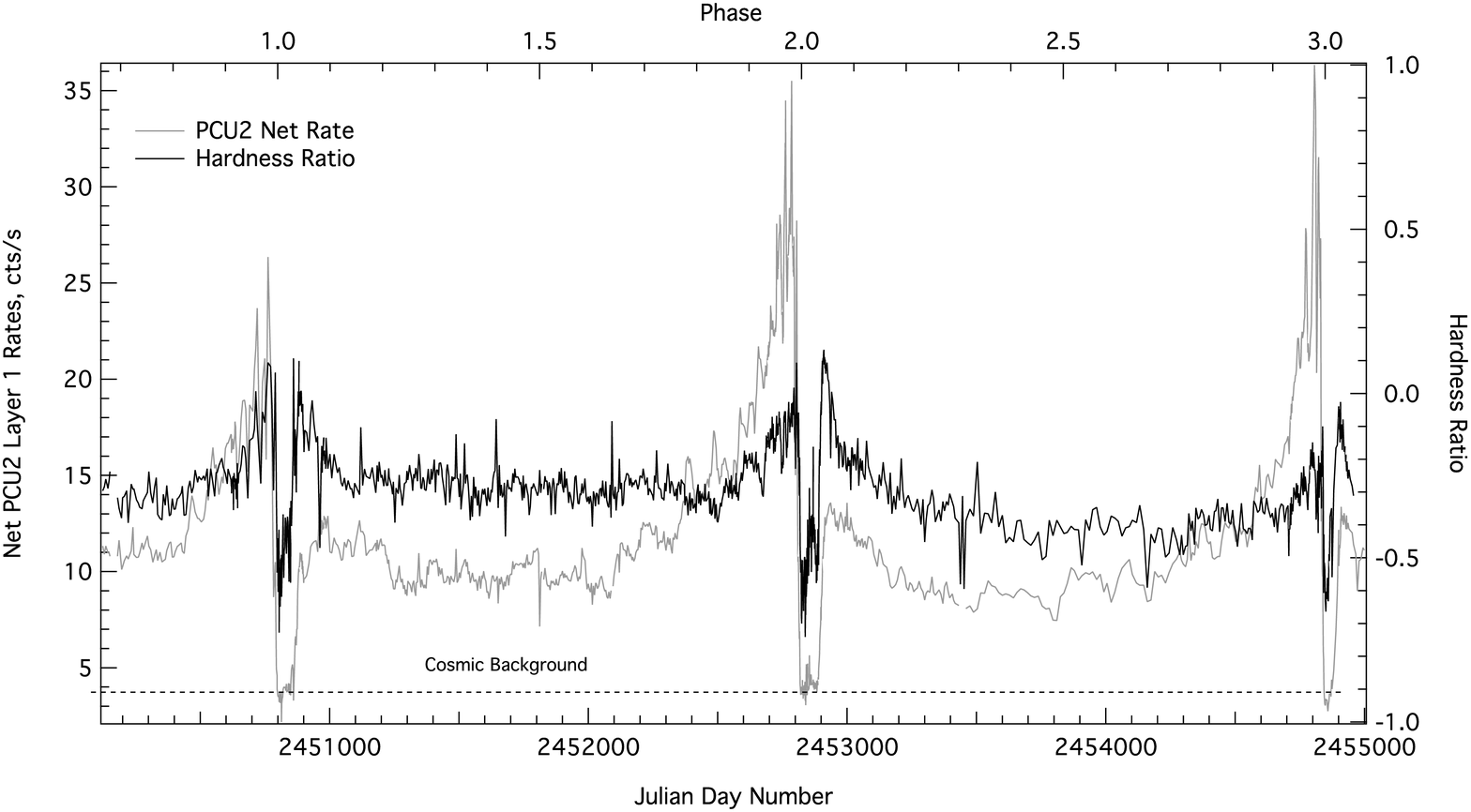} 
 \caption{Left: Overplot of Eta Car's 3 X-ray minima observed by RXTE (in the 2--10 keV band). The 2009 minimum showed an abrupt recovery compared to the two earlier minima.  Right: RXTE hardness ratio compared to the PCU2 net rate.  All three minima show a marked increase in hardness towards the end of the X-ray minima through flux recovery.}
\end{center}
\end{figure}
X-ray photometry in the 2--10 keV band of the the supermassive binary star Eta Car has been measured with the Rossi X-ray Timing Explorer from 1996--2009 (see Fig. 1).  The ingress to X-ray minimum is consistent with a period of 2024 days. The 2009 X-ray minimum began on January 16 2009 and showed an unexpectedly abrupt recovery starting after 12 Feb 2009.
This is about one month earlier than the flux recovery in the two earlier minima (in 2003.5 and 1998). This recovery roughly corresponds in phase to the ``shallow minimum'' of  Hamaguchi et al (2007 ApJ 663, 522), and suggests that for the most recent cycle the ``shallow minimum'' was very shallow indeed, or did not occur at all.  Figure 1 also shows the hardness ratio measured by RXTE compared to the RXTE fluxes.  The X-ray colors become harder about half-way through all three minima and continue until flux recovery.  The behavior of the fluxes and X-ray colors for the most recent X-ray minimum (which corresponds to the time of periastron passage of an unseen companion star) suggests a significant change in the inner wind of Eta Car and might suggest that the star is entering a new unstable phase of variable mass loss.  
\subsubsection{Optical photometry of the 2009.0 event of Eta \ Car (Eduardo Fernandez-Lajus,
Cecilia Farina,
Juan P. Calderon,
Martan A. Schwartz,
 Nicolas E. Salerno,
Carolina von Essen,
Andrea F. Torres,
Federico N. Giudici,
Federico A. Bareilles,
M. Cecilia Scalia
\&
Cintia S. Peri)}
\begin{figure}[ht]
\includegraphics[width=5.3in,angle=0]{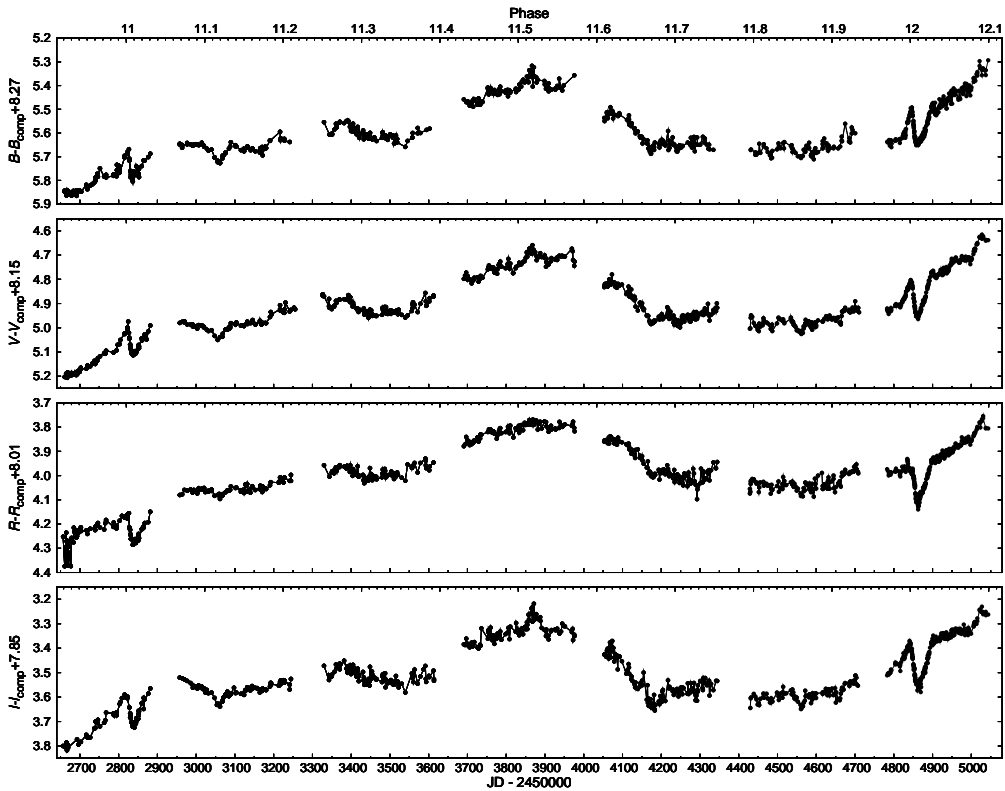}
  \caption{Optical BVRI light curves from monitoring Eta \ Car by the La Plata Observatory. While the fluxes are trending brighter, most noticeable are the broad bump associable with apastron and the two narrow drops associated with the 2003.5 and 2009.0 periastron events.}
\end{figure}
During the last ``event'' that ocurred in 2009.0, Eta  Car was the target of several
observing programs. Through our optical photometric monitoring campaign, we recorded in
detail the behavior of the associated ``eclipse-like'' event, which happened fairly on
schedule. In this work we present the resulting $UBVRI$ and H$\alpha$ light curves, 
and a new determination of the present period length.\\
\indent Our ground-based photometry was performed from the beginning of the
2009 observing season of Eta  Car, using two telescopes 
at La Plata Observatory and Complejo Astron\'omico El Leoncito, 
both located in Argentina.
The $UBVRI$ and H$\alpha$ light curves obtained display once more
an ``eclipse-like'' appearance. This feature is preceded by an ascending
branch which peaks a maximum one month later. A sudden drop of 
0.15 - 0.26 mag (depending on the band)  reaches a 
minimum nearly simultaneously in the six bands.
Then, the recovery phase starts and the brightness increases 
steeply up to the end of the season.
The color indices show some particularities during the event,
specially a blueing peak in $V-R$.
Although the general trend of this event is quite similar to that
of the 2003.5, there are some differences, specially the
deeper dips of the minima and the high increasing rate after
the ``eclipse-like'' feature.
Our long term photometry shows some evidence of systematic
brightenings of the central region (r $<$ 3'') relative to the
complete ``Homunculus'' (r $<$ 12'') occurring just after each
of these last two events.\\
\indent Our results provided more observational evidence on the periodic 
origin of the events occurring at Eta  Car,  in accordance with 
the proposed binary nature of this object.
 \vfill\eject\subsubsection{VLTI/AMBER interferometry and VLT/CRIRES spectroscopy of Eta Car across the 2009.0 spectroscopic event (Gerd~Weigelt, Jos\'e H.~Groh,  Thomas~Driebe, Karl-Heinz Hofmann, Stefan Kraus, Dieter Schertl, P. Bristol, Augusto Damineli, Theodore Gull, Henrik Hartman, Florian Kerber, Florentin Millour, Koji Murakawa \& Krister E. Nielsen)} 
Eta  Car's 2009.0 spectroscopic event provided a unique opportunity to study the changes of Eta  Car's primary wind and wind-wind
interaction region. The goals of  VLTI/AMBER observations in 2008 and 2009 were to study the wavelength-dependent shape of Eta 
Car's aspherical stellar wind and wind-wind interaction region
across the 2009.0 spectroscopic event. We carried out a large number of VLTI/AMBER observations with spectral resolution of 12000
in April 2008, January 2009, March 2009, and April 2009. We measured that the size of the wind did not significantly change at the
wavelength of the Br$\gamma$\,2.16\,$\mu$m line during our event observations from Jan 1 to 8. However,
during the event, the size of the He\,I~2.06\,$\mu$m emitting region collapsed from 17 mas (continuum-subtracted 50\%
encircled energy diameter before the event) to only 6 mas during the event. Therefore, we found strong evidence for the collapse of the wind-wind interaction zone during periastron passage.\\
\indent In addition, we obtained near-IR long-slit spectroscopy of Eta  Car with very high spatial ($0.2''$) and spectral ($R$\,=\,100\,000) resolution using VLT/CRIRES.
These unique data provided definitive evidence that high-velocity material, up to $\sim-1900~{\rm km\,s^{-1}}$, was present in the wind region of Eta Car during the 2009.0 periastron passage. A broad, high-velocity absorption is seen in He {\sc I}  $\lambda$10833 only in the spectrum of 2008 Dec 26 -- 2009 January 07, which strongly suggests a connection with the periastron passage, since a brief appearance of high-velocity material was also detected during previous periastron passages. We suggest that the high-velocity absorption is either formed directly in the wind of the companion star or, most likely, is due to shocked, high-velocity material from the wind-wind collision zone.
\subsubsection{HeII 4686A in Eta Car: The Data and Modeling (Augusto Damineli, Mairan Teodoro, Joao E. Steiner,  Nidia I. Morrell,  Rodolfo H. Barba, Gladys Sollivela, Roberto C. Gamen, Eduardo Fernandez-Lajus, Federico Gonzalez, Carlos A.~O.~Torres, Jose Groh, Luciano Fraga, Claudio B. Pereira , Marcelo Borges Fernandes, Maria I.~Zevallos \& Peter~McGregor)}
\begin{figure}[ht]
\begin{center}
\includegraphics[width=0.75\textwidth]{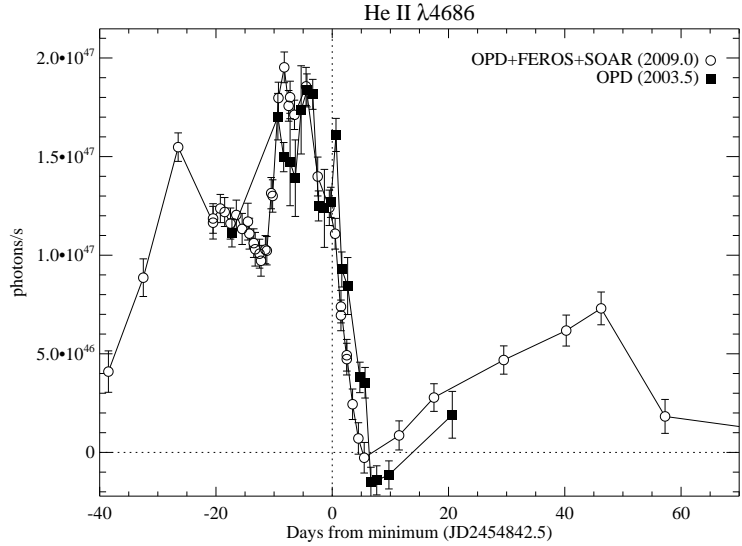}
\caption{Line flux (photons per second) in the He\,{\sc ii}~$\lambda$4686 spectral line along cycles \#11~(2003.5) and \#12 (2009.0).}
\label{fig:heii}
\end{center}
\end{figure}
The intrinsic emission of He\,{\sc ii} is quite repeatable from cycle to cycle. The He\,{\sc ii}~$\lambda$4686 line flux rises by a factor of $\approx$10 in the 2 months preceding phase zero. There are two local maxima in the month preceding the minimum and a secondary maximum $\approx$50~days after phase zero. The rising before phase zero resembles that seen in X-rays, but with remarkable differences. The He\,{\sc ii} line flux increases by a factor of $\approx$10 as compared to only a few times in X-ray emission. Both light curves collapse before phase zero, but the collapse of He\,{\sc ii} is shifted by 16.5~days relative to the X-ray collapse. The minimum in He\,{\sc ii} is reached a week after phase zero. Since the X-ray variability is measured in the range 2--10 keV, and comes mainly from the vertex of the wind-wind shock cone, it is probably not common to  the He\,{\sc ii} emitting region, which comes from gas at lower temperature. The He\,{\sc ii} line indicates a high luminosity source in the system, but it is not clear where it comes from. One possible source is the collision of the secondary star wind, since the SED derived from Parkin et al. 2009 (MNRAS 394, 1758) models indicates the presence of 10 times more He$^{+}$ ionizing photons than those passing through this atomic transition. Recombination of the shocked secondary wind is not the only source for the He$^{+}$ ionizing photons. As the shock cone migrates deep in the wind of the primary star, a huge amount of hard photons are free to escape and ionize the inner walls of the wind-wind collision zone. 
\subsection{3-D Modeling and Application}
\subsubsection{3-D models of the colliding winds in Eta Car (Julian M. Pittard, E. Ross Parkin, Michael F. Corcoran,
 Kenji Hamaguchi \& Ian R. Stevens)}
\indent A 5.5\,yr periodicity is now firmly established for Eta \ Car, with
variations seen at radio, sub-mm, infrared, optical, and X-ray
energies (Duncan \& White 2003 (MNRAS 338, 425), Abraham et al. 2005 (MNRAS 364, 922), Corcoran 2005 (AJ 129, 2018), Damineli  et al. 2008 (MNRAS 384, 1649)).
The overwhelming consensus is that this emission is regulated by the
presence of an (unseen) companion, with the emission either
originating in the wind-wind collision region between the stars (e.g.,
as for the X-rays, see Pittard \& Corcoran 2002 (A\&A 383, 636)), or being influenced by
its presence and the low-density cavity which the wind of the
companion star bores into the dense wind of the LBV primary (e.g., as
for the radio emission).\\ 
\indent The X-ray emission from Eta ~Car is believed to originate in the
hot plasma created by the high-speed wind of the companion star
shocking against the denser LBV wind (e.g.,
Pittard et al. 1998 (MNRAS 299, L5),Pittard \& Corcoran 2002 (A\&A 383, 636)). We present a recent analysis of the
{\it RXTE} X-ray lightcurve, using a 3-D model with spatially and
energy dependent X-ray emission (Parkin et al. 2009 (MNRAS 394, 1758)). The model fails
to obtain a good match to the data through the minimum and
overpredicts the hardness of {\it XMM-Newton} spectra
(Hamaguchi et al. 2007 (ApJ 663, 522)). We find that the pre-shock speed of the
companion wind must substantially decrease around periastron passage,
and that this reduction lasts for longer than expected
post-periastron. This implies that the companion wind no longer shocks
at high speed against the LBV wind at this time. We speculate that
this is either because the wind-wind collision region deforms into a
multitude of oblique, radiative shocks, or the LBV wind completely
overwhelms it and accretes onto the companion star (Soker 2005 (ApJ 635, 540)).
We conclude by presenting 3-D hydrodynamical
models of the colliding winds, noting several interesting features
as the stars swing through periastron passage.
\subsubsection{3-D Numerical Simulations of Colliding Winds in Eta Car \& WR140 (Atsuo T. Okazaki, Stanley P. Owocki, Christopher M. P. Russell, 
Thomas I. Madura \& Michael F. Corcoran)}
We report on the results from 3-D SPH simulations of colliding winds
in the supermassive binary, Eta ~Car, and the proto-typical
Wolf-Rayet binary, WR~140. For simplicity, both winds are assumed to be
either isothermal or adiabatic. Our simulations show that in
Eta ~Car the lower-density, faster wind from the secondary carves
out a spiral cavity in the higher-density, slower wind from the
primary, whereas in WR~140 it is the lower-density, primary (O4-5V)
wind that carves out a spiral cavity in the denser wind from the
secondary (WC7). Because of their very-high orbital eccentricities,
both systems show a similar, strongly asymmetric interaction surface:
the cavities are very thin on the periastron side, whereas on the
apastron side they occupy a large volume separated by thin dense
shells. A closer look, however, reveals differences caused by the
differences in  the wind momentum ratio and the speed of the slower
wind: the shock opening angle is wider and the spiral structure is
more tightly wound in Eta  Car than in WR~140. These differences are
likely to affect the observational appearance of these systems.
\subsubsection{ Precession and Nutation in Eta Car (Zulema Abraham \&  Diego Falceta-Goncalves)} 
Although the overall shape of the X-ray light curve of Eta \ Car can be explained 
by the high eccentricity of the binary orbit, other features, like the asymmetry near 
periastron passage and the short quasi-periodic oscillations seen at those epochs, have 
not yet been accounted for. We explain these features assuming that the rotation axis 
of Eta \ Car is not perpendicular to the orbital plane of the binary system. As a 
consequence, the companion star will face Eta \ Car on the orbital plane at different 
latitudes for different orbital phases and, since both the mass loss rate and the wind 
velocity are latitude dependent, they would produce the observed asymmetries in the 
X-ray flux. We were able to reproduce the main features of the X-ray light curve 
assuming that the rotation axis of Eta \ Car forms an angle of 29 degrees with the 
axis of the binary orbit. We also explained the short quasi-periodic oscillations by 
assuming nutation of the rotation axis, with amplitude of about 5 degrees and period 
of about 22 days. The nutation parameters, as well as the precession of the apsis, with 
a period of about 274 years, are consistent with what is expected from the torques 
induced by the companion star. 
 \subsubsection{Accretion onto the Companion of Eta Car (Amit Kashi \& Noam Soker)}The Accretion Model was introduced to explain observations along the
entire orbit, mainly those close around the spectroscopic event. We use
the standard parameters of the system and show that near periastron the
secondary is very likely to accrete mass from the slow dense wind blown
by the primary. The condition for accretion (that the accretion radius
is large) lasts for several weeks. The exact duration of the accretion
phase is sensitive to the winds' properties that can vary from cycle to
cycle.\\
\indent We find that: (1) The secondary accretes $\sim 2 \times 10^{-6} \rm
M_\odot yr^{-1}$ close to periastron. (2) This mass possesses enough
angular momentum to form a geometrically thick accretion belt, around
the secondary. (3) The viscous time is too long for the establishment
of equilibrium, and the belt must dissipate as its mass is 
blown in the re-established secondary wind. This processe requires
about half a year, which we identify with the recovery phase of Eta 
Car from the spectroscopic event.\\
\indent We attribute the early exit in the 2009 event to the primary wind that
we assume was somewhat faster and of lower mass loss rate than during
the two previous X-ray minima. This results in a much lower mass
accretion rate during the X-ray minimum, and consequently faster
recovery of the secondary wind and the conical shell.\\
\indent Mass transfer is an important process in the evolution of  close
massive star binaries. The high luminosity and ejected mass of many
eruptive events can be explained by mass transfer, e.g., the Great
Eruption of Eta \ Car.
\subsubsection{The outer interacting winds of Eta Car revealed by HST/STIS (Theodore R. Gull -- presented by Michael F. Corcoran)}
High spatial resolution (0.1") with moderate spectral resolution has been applied to mapping
the extended wind structure of Eta \ Car. Emission lines of [Ne\,{\sc iii}], [Ar\,{\sc iii}]. [Fe\,{\sc iii}], [S\,{\sc iii}] and  [N\,{\sc ii}] show an extended outer structure associable 
with the extended wind interaction regions. [Fe\,{\sc ii}] reveals the structure of the primary 
wind. We followed the spectro-images of these lines from the 1998.0 through the 2003.5 minima, finding changes in structure and velocity as the two massive winds, originating from a highly eccentric massive binary, interact.\\\indent Comparison of the forbidden line emission spatial structures to 3-D models (see Gull et al.,  2009, MNRAS 396, 1308) shows 1) that the He\,{\sc i} and H\,{\sc i}, consistent with the observations of Weigelt et al (2007, A\&A  474, 87), originate deep within the 0.1" limit of HST angular resolution, 2) that the broad [Ne\,{\sc iii}], [Fe\,{\sc iii}], [Ar\,{\sc iii}], [S\,{\sc iii}] and [N\,{\sc ii}] profiles are blue-shifted relative to the broad H\,{\sc i}, Fe\,{\sc ii} and  [Fe\,{\sc ii}] profiles. Moreover, the  spatial distributions of the high excitation, forbidden emissions are oriented in a NE to SW distribution in the form of arcuate velocity loops that evolve in strength and spatial location across the broad high state of the binary system.\\
\indent Based upon the 3-D SPH models of Okazaki (see above), the forbidden high excitation emissions originate from compressed structures in the outer regions of the interacting winds that flowed out in the previous cycle. FUV radiation is channeled by the  spiral cavity carved out by the lesser wind of Eta \ Car B, the less massive, but hotter companion, with a spectral distribution of a mid O-star. As Eta \ Car B, in the highly eccentric orbit, spends the majority of the orbit near apastron, the blue-shifted, spatial distributions of the high excitation, forbidden emission, and the excitation of the blue-shifted Weigelt condensations, demonstrate that apastron is on the near side of Eta \ Car A with periastron passing on the far side. Moreover, because of the high eccentricity of the binary system, the outer, hot, low density cavity is spirally-shifted in the orbital plane by about 45 to 60$^o$ relative to the orbital major axis, known from the X-ray curve to be tilted at 45$^o$ from the sky. Combining this information leads to placement of the orbital plane close to, if not in, the plane defined by the skirt of the Homunculus, whose planar axis is aligned to the axis of symmetry of the bipolar Homunculus and Little Homunculus.\\\indent Continued mapping of the spatial distribution provides the potential to map portions of the interacting winds as they distort throughout the 5.5 year period. 
\subsection{Mass loss in single and binary massive stars} 
\subsubsection{What causes the X-ray flares in Eta Carinae? (Anthony F. J. Moffat \& Michael F. Corcoran)} 
We examine the rapid variations in X-ray brightness (``flares''), plausibly assumed to arise in the hard X-ray emitting wind-wind collision zone (WWCZ) between the two stars in eta Car, as seen during the past three orbital cycles by RXTE.  The observed flares tend to be shorter in duration and more frequent as periastron is approached (see the figure), although the largest flares tend to be roughly constant in strength at all phases. Among the plausible scenarios (1. the largest of multi-scale stochastic wind clumps from the LBV component entering and compressing the hard X-ray emitting WWCZ,  2. large-scale corotating interacting regions (CIR) in the LBV wind sweeping across the WWCZ, or 3. instabilities intrinsic to the WWCZ), the first one appears to be most consistent with the observations.  This requires homologously expanding clumps as they propagate outward in the LBV wind and a turbulence-like power-law distribution of clumps, decreasing in number towards larger sizes, as seen in Wolf-Rayet winds.
\begin{figure}[ht]
\includegraphics[width=5.2in,angle=0]{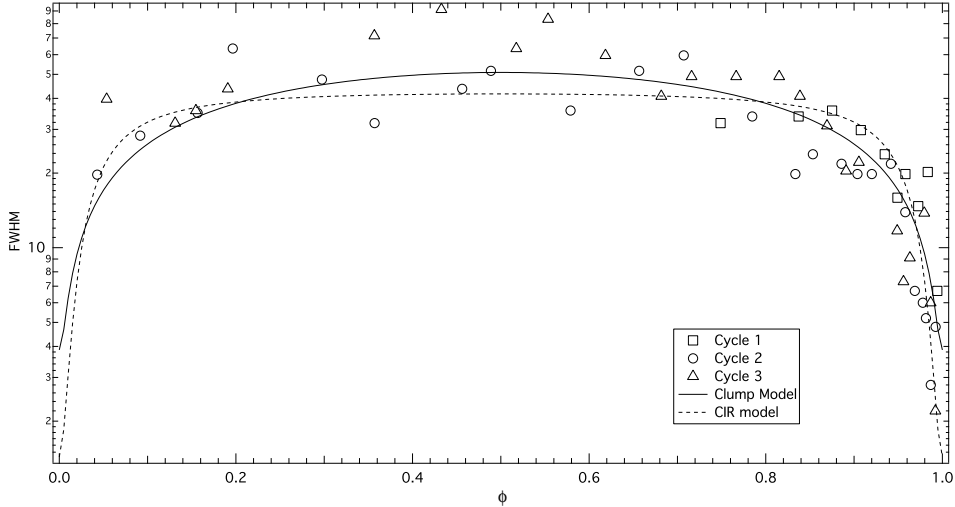}
\caption{Full width half maximum (in days) for the identified flares vs. orbital phase. Green symbols are from cycle 1, blue symbols cycle 2, and red symbols cycle 3. The smooth curves are the best-fit models: clump model, long-dashed line; CIR model, short dashed line.}
\label{fig:fwhm}
\end{figure}
\subsubsection{Revealing the mechanism of the Deep X-ray Minimum of Eta  Car
(Kenji Hamaguchi, Michael F. Corcoran \& the Eta Car 2009 Campaign Observational Team)}
The multi-wavelength observing campaign of the colliding wind binary system, Eta  Car,
targeted at its periastron passage in 2003 presented a detailed view of the flux and
spectral variations of the X-ray minimum phase.
The X-ray spectra showed a strange Fe K line profile, without significantly varying the
hard band slope above 7~keV.
The result, combined with 3-D modeling studies, suggests that the X-ray minimum originates
from either an eclipse of most of the emission by a porous absorber
or a large change of the plasma emissivity.\\
\indent The key to solve this problem would be in the deep X-ray minimum phase when X-ray emission
from the central point source plunges.
We therefore launched another focussed observing campaign of Eta  Car 
with the {\it Chandra}, {\it XMM-Newton} and {\it Suzaku}
observatories during the periastron passage in early 2009.
Five {\it Chandra} spectra taken during the deep minimum revealed 
an underlying non-variable X-ray component from the central point source.
With similar X-ray characteristics,
it would be the Central Constant Emission (CCE) component discovered in 2003. Instead, the 2009 data showed it has a very hot plasma of {\it kT} $\sim$4$-$6~keV.
The other, variable component, probably originating in the wind-wind collision (WWC),
decreased from the hard energy band above $\sim$4~keV around the onset of the deep minimum
and recovered only in the hard band at the end.
These phenomena are consistent with a picture that the hottest plasma at the WWC convex
was hidden behind an optically thick absorber first and cooler plasmas in the WWC tail followed: 
i.e., the deep minimum would be driven by an X-ray eclipse.
On the other hand, {\it Suzaku} did not find any extremely embedded X-ray source ($N_{\rm H} \lesssim$ 10$^{25}$~$cm^{-2}$)
in spectra above 10~keV during the X-ray minimum;
{\it XMM-Newton} spectra showed strong deformation in the iron K line as in the last cycle;
the X-ray minimum recovered earlier in 2009 without significant $N_{\rm H}$ change from the 2003 cycle.
These results suggest that the WWC plasma activity significantly changed during the X-ray minimum.
\subsection{LBVs, Massive Binaries and SNs: Is there a Connection?}
\subsubsection{Connections between LBVs and Supernovae (Nathan Smith)}
I will discuss the properties of LBV eruptions inferred from their circumstellar nebulae 
and from their light curves in historical examples and extragalactic Eta Carinae analogs. 
Recent observations of supernovae, especially those of the Type IIn class, suggest that 
these supernovae undergo precursor outbursts with masses, velocities, kinetic energies, 
and composition similar to the 1843 giant eruption of Eta Carinae and non-terminal 
giant eruptions of other LBVs. This possible connection offers valuable clues to the final pre-SN evolution of massive stars that contradict current paradigms, and it emphasizes that giant LBV eruptions (or events like them) represent a key long-standing 
mystery in astrophysics that begs for our attention. 
\subsubsection{The S-Dor phenomenon in Luminous Blue Variables (Jose H. Groh)}
While Luminous Blue Variables (LBVs) have been classically thought to be rapidly
evolving massive stars in the transitory phase from O-type to Wolf-Rayet stars,
recent works have suggested that LBVs might surprisingly explode as a
core-collapse supernova. Such a striking result highlights that the evolution of
massive stars through the LBV phase is far from being understood. LBVs exhibit
photometric, spectroscopic, and polarimetric variability on timescales from days
to decades, probably caused by different physical mechanisms. \\
\indent I presented the latest results on the long-term S Dor-type variability of LBVs, in particular
regarding changes in bolometric luminosity, the Humphreys-Davidson limit, and the role of rotation. The S Dor-type variability characterized by irregular visual magnitude changes on timescales of decades, with a typical amplitude of $\Delta V \simeq1-2$ mag, and corresponding changes in effective temperature and hydrostatic radius. During visual minimum, the star is typically hot, while at visual maximum, a cooler effective temperature is obtained. How the S Dor-type variability relates to the powerful giant eruptions is not clear, although it could be possible that a relatively large amount of stellar mass, which is not ejected from the star, is taking part in the S Dor-type variability. This would suggest that the S Dor-type variability is a failed giant eruption.\\ \indent At least for AG Car, a significant reduction ($\sim50\%$) in the inferred bolometric luminosity from visual minimum to maximum has been determined, and a high rotational velocity has been obtained during minimum. I will  present evidence that fast rotation is typical in Galactic LBVs that show S-Dor type variability, and will discuss how these recent results put strong constraints on the progenitor, current evolutionary stage,  and fate of LBVs.
\vfill\eject
\subsubsection{Pulsational instability in massive stars: implications for SN and LBV progenitors (Matteo Cantiello \& Sung-Chul Yoon)} Most massive stars experience a pulsational instability induced by $\ukappa-$mechanism, when 
the surface temperature sufficiently decreases. The amplitude of pulsations grows very 
fast, and may result in very high mass loss rates. We propose a new scenario for mas- 
sive star evolution based on our new calculations of this pulsational instability, where 
the initial mass of SNe progenitors increases according to the order: 
SN IIp$-->$ SN IIn$-->$SN IIL$-->$SN IIb$-->$SN Ib/c. 
Moreover, the pulsation appears strong in the early core He-burning stage for M 
$\ge$40M$_o$, 
and may lead to the formation of LBVs. We also argue that stellar eruptions like 
SN 2008S may be related to this instability. 
\subsubsection{Hydrodynamical Models of Type II-P SN Light Curves (Melina C. Bersten, Omar Benvenuto, \& Mario Hamuy)}
\begin{figure}[h!]
\hspace{0cm} \includegraphics[width=4.5cm,
  angle=-90]{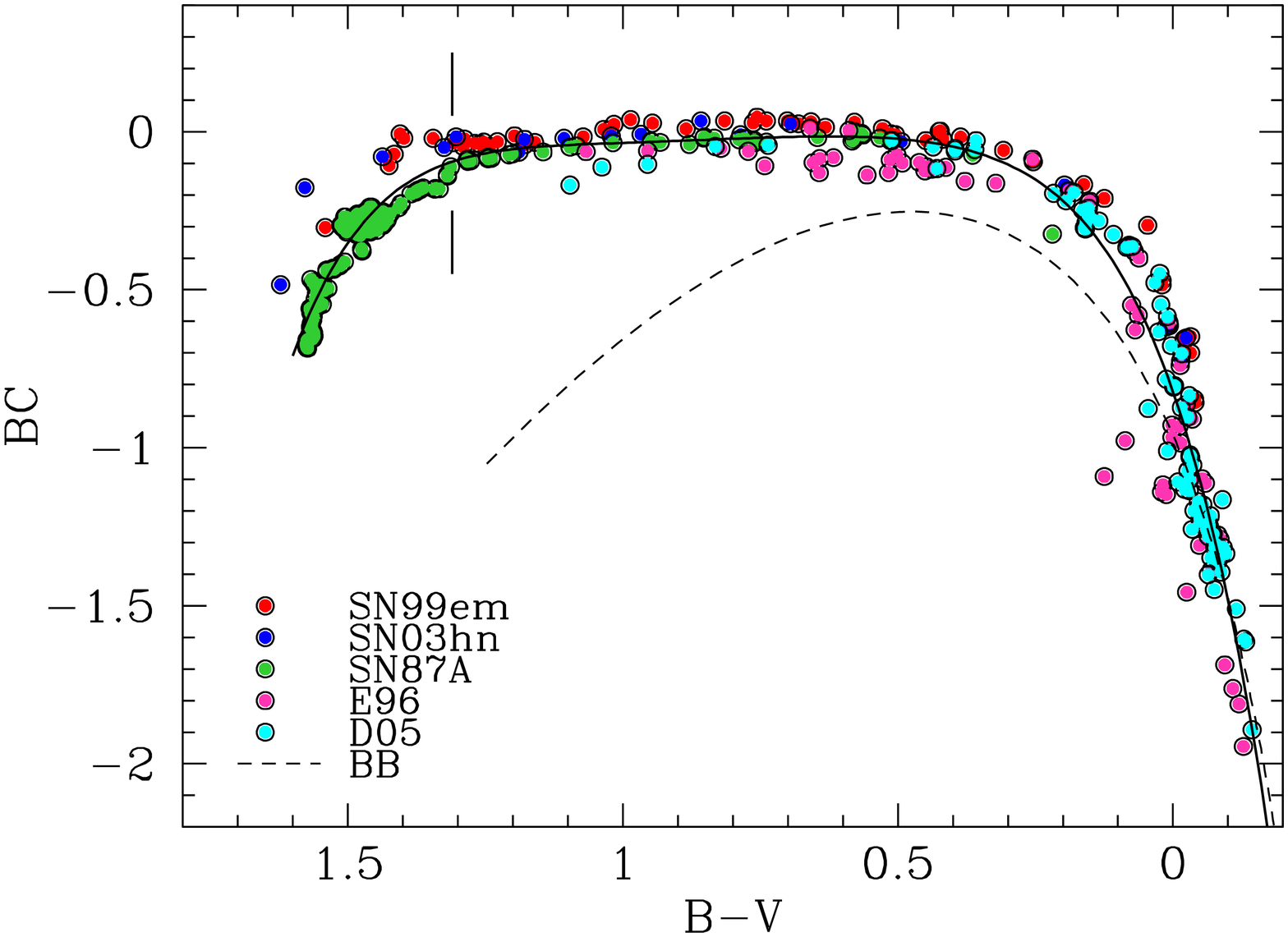} \hspace{1cm} 
\includegraphics[width=4.5cm, angle=-90]{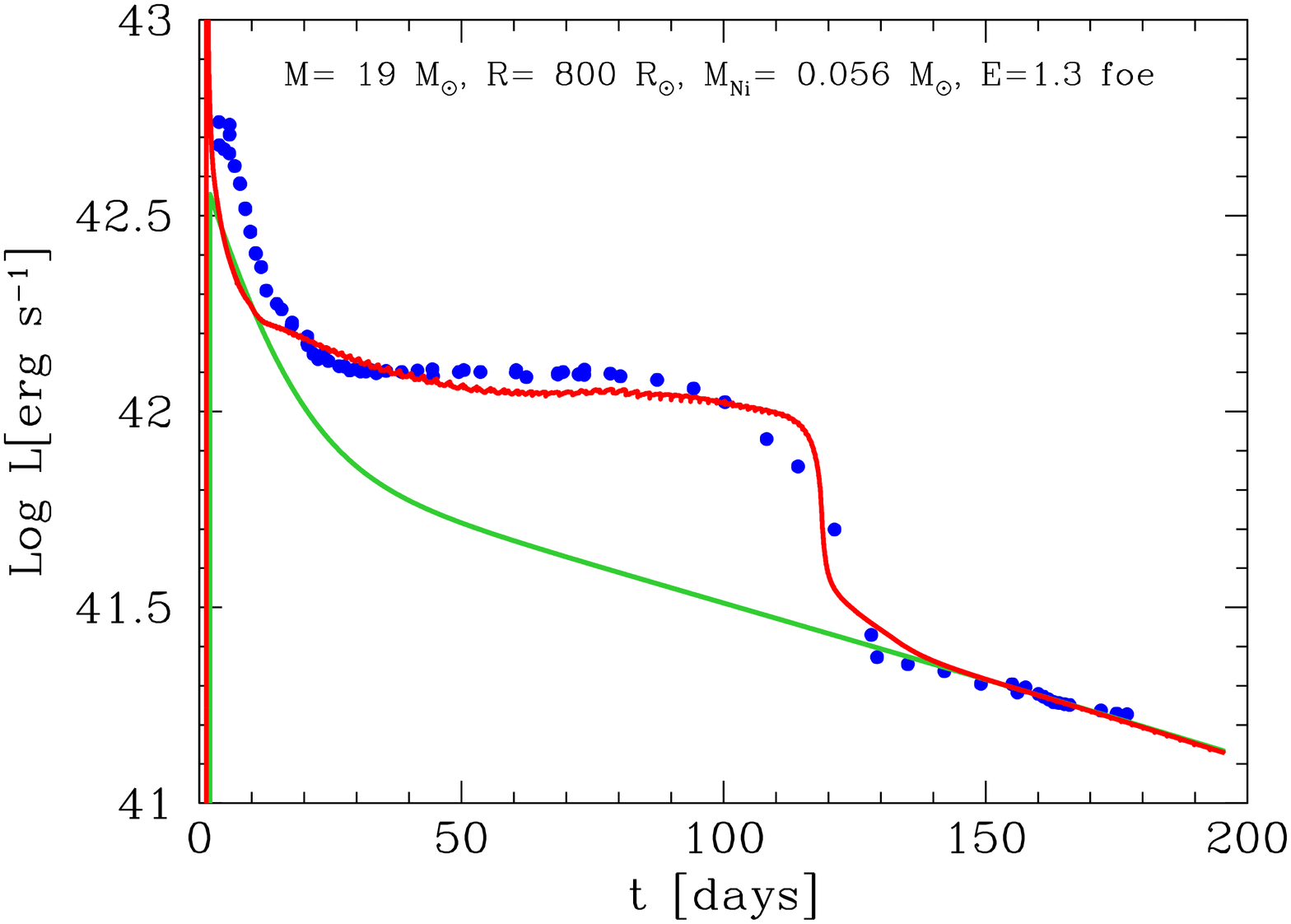}
\label{fig:lum}\caption {Hydrodynamical models for TypeII-P SN Light Curves. Left: Bolometric correction versus B-V. Right: Light curve for SNeII-P}
\end{figure}
We present computations of bolometric light curves
(LC) of type II plateau supernovae
(SNe~II-P) obtained using a newly developed,
one-dimensional Lagrangian hydrodynamic code with flux-limited
radiation diffusion. We derive a calibration for bolometric
corrections (BC) from $BVI$ photometry (see figure \ref{fig:lum}, left) with
the goal of comparing 
our models with a large database of high-quality $BVI$ light curves of
SNe~II-P. The typical scatter of our calibration is 0.1 mag.
As a first step, in our comparison we have determined the
physical parameters (mass, radius and energy) of two very well observed
supernovae,  SN~1999em (see figure \ref{fig:lum}, right) and SN~1987A. Despite 
the simplifications used in our code we obtain a remarkably good
agreement with the observations and the parameters 
 derived are in excellent concordance with previous studies of
 these objects.
\subsection{Massive Binaries and Eta Car: What is the Relationship?}
\subsubsection{WR140 \& WR25 in X-ray relation to Eta Car (Andrew M. Pollock \& Michael F. Corcoran)} 
WR~25 \hbox{(WN6ha+O)} and WR~140 \hbox{(WC7+O5)} are both X-ray bright
binaries of long period and high eccentricity, whose individual stellar and wind and collective
binary parameters are much better known than those of Eta ~Car.
Observations at
different orbital phases thus show how X-rays are produced by colliding
winds under physical
and geometrical conditions that are quite well defined at any one time but which
vary considerably around the orbit.
As WR~25 is 7'~from Eta ~Car, there are more observations than
would otherwise be the case, a few of which during the 2003 \textit{XMM--Newton}\ campaign
led to the recognition of brightness and absorption
variations that were soon shown to coincide with a periastron passage of the
208-day $e\approx0.6$ optical radial velocity orbit discovered by Gamen et al. 2006 (\textit{A\&A} 460, 777).
Their orbit was used in early 2008 to plan a month-long daily ToO campaign
with the soft X-ray XRT instrument aboard the \textit{Swift}\ GRB Observatory.
As well as the relatively shallow eclipse by the extended Wolf-Rayet wind, a sudden overall decrease
between quadrature and conjunction is most obviously interpreted as a stellar
eclipse by the WN6ha primary, thought to be one of the most massive stars in
the Galaxy. Repeatability is good within the relatively modest
statistical limits of the few dozen measurements available,
spread unevenly over several cycles.
The luminosity increases monotonically between apastron and periastron
from the surface that provides the backdrop for
the eclipses.\\\indent
Observing conditions for WR~140
are more favourable. It has an orbit well-established by Marchenko et al. 2003 (\textit{ApJ}, 596, 1295)
of  longer 7.94-year period and higher $e\approx0.881$ eccentricity. It is also a
brighter X-ray source.
As a result, measurements are more precise and the phase density much higher.
Weekly hard X-ray monitoring with \textit{RXTE}\ started just before the 2001
periastron passage, increasing to daily measurements in the approach to the 2009
periastron with recent measurements also made with \textit{Swift}, \textit{Suzaku}\ and \textit{XMM--Newton}.
Preliminary analysis of the \textit{RXTE}\ data show the
same general type of eclipse events seen in WR~25 but in greater detail and with
significant differences. For example, the luminosity maximum apparently occurs a few weeks before
periastron and even before conjunction. with asymmetries before and after periastron.
The adiabatic $1/D$ luminosity law gives a poor description throughout the orbit and there
were no obvious flares like those seen in Eta ~Car.
High resolution \textit{Chandra}~data obtained at 4 phases show very small changes in
shape between apastron and O-star conjunction in a spectrum dominated, perhaps surprisingly given the expected collisionless
nature of the shocks concerned, by a smooth continuum probably from hot electrons.
The lines imply complete mixing of shocked material from both winds.
Details of the velocity profiles are more difficult to understand, especially the absence of the highest
velocity blue-shifted material near periastron.
\subsubsection{The Erupting Wolf-Rayet System HD 5980 in the SMC: A (Missing) Link in Massive Stellar Evolution or a Freak?
(Rodolfo H. Barba)}The Wolf-Rayet eclipsing binary system HD 5980 in the Small Magellanic Cloud has 
shown a peculiar behaviour along the past years. In 1994 the star developed an un- 
predicted eruption and changed its spectrum from WN-type to one resembling those 
of Luminous Blue variables (LBV). In this presentation, I will review observational 
aspects of this unique system, emphasizing those similarities and differences with extreme LBV objects like Eta \ Car. I will briefly describe a century of photometric and 
spectroscopic records of the star, and depict a new analysis of the spectroscopic data 
obtained during the outburst phase, and the present WN-E stage. Also, I will discuss 
the different scenarios proposed to explain the LBV-like behaviour (rapid rotators, tidal 
interactions, single star evolution).
\subsubsection{The Extragalactic Eta Car Analogs (Schuyler D. Van Dyk)}Powerful eruptions of massive stars, such as Eta  Car are often referred to as
``supernova (SN) impostors,'' because some observational aspects can mimic the
appearance of a true SN.  During the Great Eruption during the 1800's
of Eta  Car, the star greatly exceeded the Eddington limit, with its
bolometric luminosity increasing by $\sim$2 mag.  The total luminous output of such an eruption ($\sim 10^{49.7}$ erg) can rival that
of a SN, to such a degree that some impostors initially are assigned 
designations as SNe, even in modern extragalactic SN searches.  A number of extragalactic SN impostors are known, such
as SNe 1954J, 1961V, 1997bs, 1999bw, 2000ch, 2001ac, 2002kg, 2003gm, 
NGC 2363-V1, etc.  I will present here the latest results for those that can be considered Eta  Car analogs.  
Not all impostors are as powerful as Eta  Car, and are therefore not considered true analogs to Eta  Car; some cases are more like the 
``classical'' LBVs (e.g., S~Dor), where the bolometric luminosity remains constant during an eruption, as the star's envelope expands or its wind becomes optically thick, and the apparent temperature cools to $\sim$8000 K.
Like Eta ~Car, the precursor star for each analog is expected to survive the eruption and return to relative quiescence.  
Some have had eruption survivors identified (SNe 1954J, 1961V), using the {\sl Hubble Space Telescope}, some have seemingly
"vanished" after outburst (SNe 1997bs, 1999bw), and one (SN 2000ch) continues in outburst after almost a decade. Only one (SN 1999bw) has shown evidence for dust emission, based on {\sl Spitzer Space Telescope\/} observations, and the emission has apparently faded from detection.
Studying the characteristics of the analogs provides us with a greater understanding of Eta  Car itself and of the evolution of very massive stars.
\subsection{Summary and Discussion (Nidia I. Morrell, Michael F. Corcoran, Anthony F.Moffat \& Julian Pittard)}After a brief brainstorm session, Mike Corcoran, Tony Moffat, Nidia Morrell and Julian Pittard came up with the following list of questions and highlights, which served as a basis for a half-hour open discussion  on  future studies of Eta \ Car:\\
\indent How to better constrain the orbital and wind parameters of both stars in Eta \ Car?\\
\indent  What is its future evolution?\\
\indent What caused the Great Eruption?  Which star erupted?
\\\indent What is the nature of the companion star? (Very urgent!)
\\\indent WhatÕs the connection between WRs, LBVs and supernovae?
\\\indent How to explain the strictly cyclic, bizarre behavior of the He II 4686 emission, which emerges only within several months of periastron passage?
\\\indent What is the role of a companion star in driving the formation, evolution and instabilities of Eta \ Car and other binary LBVs?
\\\indent Does dust form in Eta \ Car?
\\\indent Does Eta \ Car pulsate?
\section{Posters} 
\subsection{A full cycle 7 mm light-curve of Eta Car (Zulema Abraham, Pedro P.  Beaklini \& Carlo Miceli)}
It is now well established that the light curve of Eta Carinae has a periodic behavior 
at all wavelengths, from mm waves to X-rays. These light curves are characterized by 
the presence of a sharp dip, with duration that depends on wavelength, being longer 
at X-rays. At mm wavelengths, the dip was detected during the last four cycles, but 
only during the 2003.5 minimum the light curve was obtained with daily resolution. 
At that epoch, the 7 mm light curve, obtained with the Itapetinga radiotelescope, in 
Atibaia, Brazil, followed the X-ray decaying behavior but showed a strong peak, not 
seen at other wavelengths, before reaching the minimum. This peak was attributed to 
free-free emission of the 107 K optically thick gas located at the wind-wind collision 
contact surface. Here, we report the 7 mm light curve of the complete 2003-2009 
cycle, including the 2003.5 and 2009.0 minima, both obtained with daily resolution. 
We show for the Þrst time that: (a) the duration of the minima are the same at 7 mm 
and at X-rays; (b) The peak at 7 mm seen after the minimum is 2003.5 appeared again 
in 2009.0, with the same phase, duration and shape; (c) two other strong peaks were 
observed before the 2009.0 minimum, coincident with the peaks observed at X-rays, 
which supports the previous assumption that they are formed at the wind-wind shock 
interface.
\subsection{The multiple zero-age main-sequence O star Herschel 36 (Julia I. Arias, Rodolfo H. Barba, Roberto C. Gamen, Nidia I.  Morrell, Jesus Maiz Apellaniz, Emilio  J. Alfaro, Nolan R.  Walborn, Alfredo Sota, Christian M. Bidin)}
We present a study of the zero-age main-sequence O star Herschel 36 in M8, based 
on high-resolution optical spectroscopic observations spanning six years. This object 
is deÞnitely a multiple system. We propose a picture of a close massive binary and 
a companion of spectral type O, most probably in wide orbit about each other. The 
components of the close pair are identiÞed as O9 V and B0.5 V. The orbital solution 
for this binary is characterized by a period of 1.5415$\pm$0.00001 days. With a spectral 
type O7.5 V, the third body is the most luminous component of the system. It also 
presents radial velocity variations with short (a few days) and long (hundreds of 
days) timescales, although no accurate temporal pattern can be discerned from the 
available data. Some possible hypotheses to explain the variability are brießy 
addressed and further observations are suggested. 
\subsection{Spatially extended wind emission in the massive binary systems VV Cep \& KQ Pup (Wendy Hagen Bauer, Theodore R. Gull, Philip  Bennett \& Jahanara Ahmad)}VV Cep and KQ Pup are binary systems consisting of M supergiant primaries with B 
main-sequence companions which orbit within the extensive M supergiant winds. VV 
Cep undergoes total eclipses and was observed with the HST/STIS Spectrograph at 
several epochs which spanned total eclipse through "chromospheric eclipse" as lines 
from ions like Fe\,{\sc i} weakened and disappeared through first quadrature. KQ Pup comes 
close to eclipsing its hot companion and was observed to be in chromospheric eclipse 
(showing weak absorption from Fe\,{\sc i} in the M supergiantÕs chromosphere) by STIS in 
October 1999. Two-dimensional reprocessing of the STIS echelle spectra has revealed 
spatially extended emission in all observations of these two systems. Emission arising 
from gas thought to be associated with the hot component shows spatial extension 
consistent with the STIS spatial point spread function. The spatially extended flux 
seen outside total eclipse arises from emission in transitions expected to be observed 
from the winds of cool supergiants. VV Cep was observed at enough epochs to map 
out radial velocity structure within the wind. It is consistent with model predictions 
for wind flow in a binary system in which the wind outflow is comparable with the M 
supergiantÕs orbital velocity. Spatially resolved wind and wind interaction structures 
of these two stars and of Eta \ Car reinforce the need for imaging spectroscopy and 
added capabilities of integral field units for mapping these complex interacting binary systems. 
\subsection{Abundances and depletion of iron-peak elements in the Strontium filament of Eta Car (Manuel A. Bautista, Henrik Hartman, Marcio Melendez, Theodore R.  Gull, Katharina Lodders  \& Mariela Martinez)}We carried out a systematic study of elemental abundances in the Strontium Filament, 
a peculiar metal-ionized structure located in the skirt plane of the Homunculus, ejecta 
surrounding Eta \ Car. To this end we interpret the emission spectrum of neutral C 
and singly ionized Al, Sc, Ti, Cr, Mn, Fe, Ni, and Sr using multilevel non-LTE models 
for each ion. The atomic data for most of these ions is limited and of varying quality, 
so we carried out ab initio calculations of radiative transition rates and electron impact 
excitation rate coefÞcients for each of these ions. The observed spectrum is consistent 
with an electron density 
$\approx10^7$cm$^{-3}$ and a temperature between 6000 and 7000 K. 
The observed spectra are consistent with large enhancements in the gas phase Sr/Ni, 
Sc/Ni, and Ti/Ni abundance ratios relative to solar values. Yet, the abundance ratios 
Cr/Ni, Mn/Ni, and Fe/Ni are roughly solar. We explore various scenarios of elemental 
depletion in the context of nitrogen-rich chemistry, given that the stellar ejecta has 
enriched nitrogen at the expense of greatly depleted oxygen and carbon due to mixing 
in the $>$60 $M_\odot$\ star. Finally, we discuss the implications of these findings for 
the generation of dust during the evolution of supermassive stars from main sequence 
to pre-supernova stage.
\subsection{A fast ray tracing disk model for 10$\mu$ interferometric data fitting: First application on the B[e] star CPD57 2874 (Philippe Bendjoya, Giles   Niccolini,  \& Amando D. de Souza)}We present here a parametric dust disk model (P2DM) developed  to fit interferometric observations in a much faster computing time than the classical Monte 
Carlo Modeling Approach. P2DM combined with a Levenberg-Markward minimisation algorithm allows us to derive  both crucial physical and geometrical parameters. This model is restricted to wavelengths around and above 10 microns (no gas, no scattering) making it useful for VLTI-MIDI (and future MATISSE) observations and implies 
that  more elaborate modelling is necessary to get a deeper understanding of the 
physical processes responsible of the observed disks. Neverthelss, this fast and physical model is useful for exploring the physical parameter phase space and to provide 
starting values for more powerful models. We  present the model and its applica- 
tion to the supergiant B[e] CPD -57 2874 star observed with VLTI-MIDI. 
\subsection{A search for relics of interstellar bubbles originated by LBV progenitors (Cristina E. Cappa, Silvina Cichowolski, Javier Vasquez \& J. R. Rizzo)}The strong stellar winds of massive O stars sweep up and compress the surrounding 
gas creating interstellar bubbles in their environs. In this modified environment, massive stars evolve into Luminous Blue Variables (LBVs), which are the 
immediate progenitors of WR stars. 
Using the Canadian Galactic Plane Survey (CGPS) and Southern Galactic Plane Survey
(SGPS) we searched for H{\sc i} interstellar bubbles associable with O-type progenitors of a number of galactic LBVs and LBV
candidates.
We found H{\sc i} cavities and shells that probably originated from the massive 
progenitors of P Cygni, G79.29+0.46, AG Carinae, and He3-519.
\subsection{Massive binaries and rotational mixing (Selma E. de Mink, Matteo  Cantiello, Norbert  Langer  \& Onno R. Pols)}In massive stars fast rotation is the cause of efficient internal
mixing, which leads to the transport of hydrogen burning products from the
core to the stellar envelope.
This results in hot and overluminous stars, which stay compact as they
gradually evolve into massive helium stars (e.g. Yoon \& Langer, 2005).
While non-rotating stars in close binaries experience severe mass
loss as soon as their radius exceeds the Roche lobe radius,
fast-rotating stars, which are efficiently mixed, stay compact and can
avoid the onset of mass transfer.\\
\indent This can occur in wide binaries (orbital periods much larger than
about 10 days) where the rotation rate of the stars is not affected by
tides during the main sequence evolution. Alternatively, this can
occur in massive binaries with orbital periods smaller than 3 days.
Tides force the stars to rotation rates high enough to trigger
efficient mixing (De Mink et al. 2008, 2009).  This type of evolution
leads naturally to the formation of compact Wolf-Rayet binaries and is
potentially interesting as an explanation for the formation of massive
black hole binaries such as M33~X-7 and IC10~X-1.
\subsection{MHD numerical simulations of wind-wind collisions in massive binary systems (Diego Falceta-Goncalves \& Zulema Abraham)}In  past years, several massive binary systems have been studied in details at both radio and X-rays wavelengths, revealing a 
whole new physics present in such systems. Large emission intensities from thermal and non-thermal sources showed us that most of 
the radiation in these wavelengths originates at the wind-wind collision region. OB and WR stars present supersonic and massive winds 
that, when under collision, emit largely in X-rays and radio due to the free-free radiation, as well as in radio due to synchrotron 
emission. However, in the latter case, magnetic fields play an important role on the emission distribution. Astrophysicists have been 
modeling free-free and synchrotron emission from massive binary systems based on purely hydrodynamical simulations and ad hoc assumptions 
regarding the distribution of magnetic energy and the field geometry in order to study the non-thermal source. In this work we provide 
a number of the first MHD numerical simulations of wind-wind collision in massive binary systems. We study the free-free emission, 
characterizing its dependence on the stellar and orbital parameters. We also study, self consistently, the evolution of the magnetic 
field at the shock interfaces, obtaining also the synchrotron energy distribution integrated along different lines of sight.
\subsection{On the peculiar variations of two southern B[e] stars (Marcelo Borges Fernandes, Michaela  Kraus, Olivier Chesneau, Jiri  Kubat, Armando Domiciano  de Souza, Francisco X.  de Araujo, Philippe Stee  \& Anthony Meilland)}In this work, we present the peculiar variations shown by two B[e] stars, namely the 
SMC supergiant LHA115-S23 and the galactic unclassified object HD50138, mainly 
based on high resolution optical spectroscopic data. The spectra of LHA115-S23 revealed the disappearance of photospheric He\,{\sc i} absorption lines in a period of only 11 
years. Due to this, the star has changed its MK classiÞcation from B8I to A1Ib, becoming the first A[e] star identified. Concerning HD50138, the brightest known B[e] star, 
based on our data, taken with a difference of 8 years, it is possible to see the presence 
of strong spectral variations, probably associated with a new outburst, which took place 
prior to 2007. A detailed spectroscopic description, the projected rotational velocities, the 
modeling of their spectral energy distributions, and the discussion about the possible 
nature and circumstellar scenarios for these two curious B[e] stars are provided.
\subsection{Interferometric analysis of peculiar stars with the B[e] phenomenon (Marcelo Borges Fernandes, Olivier  Chesneau, Denis Mourard, Michaela Kraus, Philippe  Stee, Armando Domiciano de Souza, Alex Carciofi, Florentin Millour, Anthony Meilland, Philippe Bendjoya,  Samer Kanaan, Giles Niccolini \& Olga Suarez)}Stars that present the B[e] phenomenon are known to form a very heterogeneous group. 
This group is composed by objects in different evolutionary stages, like high- and low- 
mass evolved stars, intermediate-mass pre-main sequence stars and symbiotic objects. 
However, more than 50\% of the confirmed B[e] stars have unknown evolutionary stages, being called as unclassified B[e] stars. The main problem  
is  the absence of reliable physical parameters and of knowledge of their circumstellar geometries. Based on this, high-angular resolution interferometry is certainly an important tool to answer several questions concerning the nature of these 
stars, including a possible evolutionary link between B[e] supergiants and LBV stars, 
like Eta \ Car. In this work, we present the results related to a sample of objects, 
namely HD50138, HD45677, HD62623 and MWC361 based on observations using 
VLTI/MIDI, VLTI/AMBER and CHARA/VEGA. 
\subsection{Numerical models for  19th century outbursts of Eta Car (Ricardo F.  Gonzalez Dominguez)}We present new results of two-dimensional hydrodynamical
simulations of the eruptive events of the 1840s (the great)
and the 1890s (the minor) eruptions suffered by the massive
star, Eta  Car. The two bipolar nebulae commonly known as
the Homunculus (H) and the Little Homunculus (LH) were formed
from the interaction of these eruptive events with the
underlying stellar wind. We assume a colliding wind scenario
to explain the shape and the kinematics of both Homunculi.
Adopting a more realistic parametrization of the phases
of the wind, we show that the LH is formed at the end of the
1890s eruption when the post-outburst Eta  Car wind collides
with the eruptive flow, rather than at the beginning (as
claimed in previous works; Gonz\'alez et al. 2004a, 2004b).
The regions at the edge of the LH become Rayleigh-Taylor
unstable and develop filamentary structuring that shows
some resemblance with the observed spatial structures in the
polar caps of the inner Homunculus (Smith 2005). We also find
the formation of some tenuous equatorial, high-speed features.
\subsection{Discovery of a new WNL star in Cygnus with Spitzer (Vasilii Gvaramadze, Sergei  Fabrika, Wolf-Rainer  Hamann, Olga Sholukhova, Azamat F.  Valeev, Vitaly P. Goranskij, Anatol M. Cherepashchuk, Dominik J.  Bomans \&  Lidia M. Oskinova)}
We report the serendipitous discovery of an infrared ring nebula in Cygnus using 
the archival data from the Cygnus-X Spitzer Legacy Survey and present the results of study of 
its central point source. The optical counterpart to this source was identiÞed by Dolidze (1971) 
as a possible Wolf-Rayet star. Our follow-up spectoscopic observations with the Russian 6-m 
telescope confirmed the Wolf-Rayet nature of this object and showed that it belongs to the 
WN8-9h subtype. 
\subsection{VLT-CRIRES observations of Eta Car's  Weigelt blobs \& Strontium Filament (Henrik Hartman, Jos\'e Groh, Thedore R.  Gull, Hans U. Kaufl, Florian Kerber, Vladilen Letokhov \& Krister E. Nielsen)}We have obtained Very Large Telescope-CRIRES observations of Eta \ Car, focused on the Weigelt condensations (WC) and the Strontium Filament (SrF). These are nebular regions, in the close vicinity to Eta Car, with complex emission line spectra. The two regions show, however, strikingly different physical conditions and abundances. The WC are driven by far-UV radiation from the hot companion (Eta Car B). The radiation is internally redistributed to hydrogen emission which enables exotic atomic photo processes, such as Resonance Enhanced Two-Photon Ionization (RETPI) and stimulated emission (LASER). The lines proposed for the stimulated emission are the 1.68 and 1.74 mm transitions from the c4F7/2 level in Fe\,{\sc ii} (i.e. the spectrum of Fe$^+$).\\
\indent The Strontium Filament received its name from the initial discovery of  [Sr\,{\sc ii}], lines from singly-ionized strontium. Modeling of the emission spectrum has revealed strange abundances (see separate poster by Bautista et al. at this meeting), and spectral lines with complex line profiles. The main emission component is consistent with a creation of the ejecta in the 1890s.\\
\indent We present a preliminary analysis of the ejecta in the NIR, using high spectral (R= 90,000) and spatial resolution ($\approx$0.3") spectra obtained with CRIRES in April 2007. The data allow us to study the individual ejecta in detail, at a spectroscopic phase where the effects due to Eta \ Car B's periastron passage is negligible.\\
\indent We all acknowledge the tremendous contributions by Sveneric Johansson and Vladilen Letokhov to the field of plasma physics, the understanding of the physical processes in the WC, and the final contribution with their book {\it Astrophysical Lasers} (Oxford, 2009).
\subsection{Radiative transfer Modeling of rotational modulations in the B supergiant HD 64760 (Alex Lobel \& Ronny Blomme)}
We develop parameterized models for the large-scale structured 
wind of the blue supergiant, HD~64760 (B0.5 Ib), based on best fits 
to Rotational Modulations and Discrete Absorption 
Components (DACs) observed with IUE in Si~{\sc iv} $\lambda$1400. 
The fit procedure employs the {\sc Wind3D} code with non-LTE 
radiative transfer (RT) in 3-D. We parameterize the density structure 
of the input models in wind regions (we term "Rotational Modulation Regions" 
or RMRs) that produce Rotational Modulations, and calculate the 
corresponding radial velocity field from CAK-theory for 
radiatively-driven rotating winds. We find that the Rotational 
Modulations are caused by a regular pattern of radial density 
enhancements that are almost linearly shaped across the equatorial 
wind of HD~64760. Unlike the Co-rotating Interaction Regions 
(CIRs) that warp around the star and cause DACs, the RMRs 
do not spread out with increasing distance from the star. 
The detailed RT fits show that the RMRs in HD~64760 have 
maximum density enhancements of $\sim$17 \% above the surrounding 
smooth wind density, about two times smaller than hydrodynamic models 
for CIRs. Parameterized modelling of Rotational Modulations reveals 
that nearly linear-shaped (or `spoke-like') wind regions co-exist 
with more curved CIRs in the equatorial plane of this fast rotating 
B-supergiant. We present a preliminary hydrodynamic model 
computed with Zeus3D for the RMRs, based on mechanical wave 
excitation at the stellar surface of HD~64760.
\subsection{Parameterized structured wind modelling of massive hot stars with Wind3D (Alex Lobel \& Jesus A. Toala)}
We develop a new and advanced computer code for modelling the 
physical conditions and detailed spatial structure of the 
extended winds of massive stars with three-dimensional (3-D) 
non-LTE radiation transport calculations of important diagnostic 
spectral lines. The {\sc Wind3D} radiative transfer code is 
optimized for parallel processing of advanced input models 
that adequately parameterize large-scale wind structures 
observed in these stars. Parameterized 3-D input models for 
Wind3D offer crucial advantages for high-performance transfer 
computations over ab-initio hydrodynamic input models. The 
acceleration of the input model calculations permits us to 
investigate and model a much broader range of physical (3-D) 
wind conditions with Wind3D. We apply the new parameterization 
procedure to the equatorial wind-density structure of Co-rotating 
Interaction Regions (CIRs) and calculate the wind velocity-structure 
from CAK-theory for radiatively-driven rotating winds. We use 
the parameterized CIR models in {\sc Wind3D} to compute the detailed 
evolution of Discrete Absorption Components (DACs) in Si~{\sc iv} 
UV resonance lines. The new method is very flexible and efficient 
for constraining physical properties of extended 3-D CIR wind structures 
(observed at various inclination angles) from best fits to DACs 
in massive hot stars. We compare the results with an accurate 
hydrodynamical model for the DACs of B0.5 Ib-supergiant HD~64760, 
and apply it to best fit the detailed DAC evolution observed 
with $IUE$ in B0 Iab/Ib-supergiant HD~164402.\vfill\eject
\subsection{ 3D modeling of eclipse-like events in Eta Car (Thomas I. Madura, Theodore R. Gull, Atsuo Okazaki   \& Stanley Owocki)} We discuss recent efforts to apply 3D Smoothed Particle Hydrodynamics (SPH) simulations to model the binary wind collision in Eta \ Car, with emphasis on reproducing BVRI photometric variations observed from La Plata Observatory. Photometric 
dips occurring concurrently with X-ray minima seen with RXTE provide further evidence for binarity in the system. We investigate the role of the unseen secondary star, 
focusing on two effects: 1) an occultation of the secondary by the slower, extended 
optically thick primary wind; and 2)a ÒBore-HoleÓ effect, wherein the fast wind from 
the secondary carves a cavity in the dense primary wind, allowing increased escape of 
radiation from the hotter/deeper layers of the pri mary s extended photosphere. Such 
models may provide clues on how/where light is escaping the system, the directional 
illumination of distant material (e.g., the Homunculus, the Little Homunculus, the Òpurple hazeÓ, 
Weigelt blobs, etc.) and the parameters/orientation of the binary orbit.
\subsection{The Other Very Massive Stars in the Carina Nebula as observed with HST (Jesus Maiz Apellaniz, Nolan R. Walborn, Nidia I. Morrell, Ed P. Nelan \& Virpi S. Niemela)}
We have used HST/ACS+FGS and ground-based data to study 10 WNha, O2-4 supergiant, and O3.5 main-sequence stars in the 
Carina Nebula. HD~93129~Aa+Ab is the most massive known astrometric binary. Its motion is currently being followed
with STIS spectroscopic observations planned for the fall of 2009. Previously unknown resolved components are detected:
an $\sim$8~M$_\odot$ star for HD~93162 (=WR~25) and two $\sim$1~M$_\odot$ stars for Tr~16-244. Overall, at least 8 
of the 11 most massive stars in the Carina Nebula are members of multiple systems. The NUV-to-NIR photometry has been 
processed with the new version (v3.1) of the CHORIZOS code using Geneva isochrones with ages of 1.0~Ma and 1.8~Ma. Most 
stars in our sample are found to have visual total extinctions between 1.0~and~2.2 mag but HD~93162 and Tr~16-244 are more 
extinguished. The ratio
of total to selective extinction $R_{5495}$ is found to vary between 3.0 and 4.5 and is positively correlated with
the total extinction. For a fixed age for the full sample, the Trumpler 14 stars are underluminous for their spectral 
types, hence implying a small age ($\lesssim$1~Ma) for the cluster. HD~93250 is overluminous for its spectral type,
a possible indication of an undetected (by spectroscopic, interferometric, or imaging methods) massive companion. 
The three WRs~(22,~24,~and~25) and HD~93129~Aa have evolutionary (initial) masses above 90~M$_\odot$, i.e. values 
comparable to that of Eta ~Car.
 \subsection{The High Angular Resolution Multiplicity of Massive Stars (Brian D. Mason, William I.  Hartkopf, Douglas R.  Gies, Theo A.  ten Brummelaar, Nils H. Turner, Chris D.  Farrington \& Todd J. Henry)}Conducted on NOAO 4-m telescopes in 1994, the first speckle survey of O 
stars (Mason et al.\ 1998) had success far in excess of our expectations. In
addition to the frequently cited multiplicity analysis, many of the new 
systems which were first resolved in this paper are of significant 
astrophysical importance. Now, some ten years after the original survey, we 
have re-examined all systems analyzed before. Improvements in detector 
technology allowed for detection of companions missed before as well as 
systems which may have been closer than the resolution limit in 1994. Also, 
we made a first high-resolution inspection of the additional O stars in the 
recent Galactic O Star Catalog of Ma\'{i}z-Apell\'{a}niz \& Walborn (2004). 
In these analyses we resolved four binaries not detected in 1994 due to the
enhanced detection capability of our current system or kinematic changes in 
their relative separation. We also recovered four pairs, confirming their 
original detection. In the new sample, stars are generally more distant and
fainter, decreasing the chance of detection. Despite this, eight pairs were
detected for the first time.\\
\indent In addition to many known pairs observed for testing, evaluation and 
detection characterization, we also investigated several additional samples
of interesting objects, including accessible Galactic WR stars from the 
contemporaneous speckle survey of Hartkopf et al.\ (1999), massive, hot 
stars with separations which would indicate their applicability for mass 
determinations (for fully detached O stars masses are presently known for 
only twelve pairs), and additional datasets of nearby red, white, sub and G
dwarf stars to investigate other astrophysical phenomena. In these 
observations, in addition to those enumerated above we resolved seventeen 
pairs for the first time.\\ \indent Massive stars have also been a important observing program for the CHARA 
Array. Preliminary results from Separated Fringe Packet solutions of 
interferometric binaries are also presented.
\subsection{Far-IR Spectroscopic Imaging of the ISM around Eta Car (Hiroshi Matsuo, Takaaki Arai, Tom Nitta \& Aya Kosaka)}To study interstellar material around Eta \ Car, we have performed far-infrared imaging spectroscopic observations using a Fourier transform spectrometer onboard the Japanese 
infrared satellite AKARI. We have obtained images of C\,{\sc ii}, N\,{\sc ii},  and O\,{\sc iii} covering the  
15 arcmin 
$\times$10 arcmin area
centered at Eta \ Car. The O\,{\sc iii} and C\,{\sc ii} lines were found wide-spread, but peaked toward Carinae nebulae, which gives an indication of interaction of 
ejecta and molecular clouds. The N\,{\sc ii} line is weak and only partially observed around Eta \ 
Car. Comparison with ionized hydrogen and non-thermal emission at millimeter-wave O\,{\sc iii} emission is coincident with ionized region while C\,{\sc ii} emission is peaked at 
different positions but similar to the position angle of the Homunculus nebulae, which 
may indicate that we are observing interactions of old ejecta with molecular clouds. 
\subsection{Stellar forensics with SNe \& GRBs: Deciphering the size 
\& metallicity of their massive progenitors (Maryam Modjaz)}Massive stars die violently. Their explosive demise gives rise to brilliant fireworks that 
constitute supernovae and long GRBs, and that are seen over cosmological distances. 
By interpreting their emission and probing their environment, we get insights into the 
size, make-up, mass loss history and metallicity of their massive progenitor stars that 
are situated at extragalactic distances.\\
\indent I will present extensive X-ray, optical and NIR data on SN 2008D which was dis- 
covered serendipitously with the NASA Swift satellite via its X-ray emission from 
shock breakout. It is a supernova of Type Ib, that is, a core-collapse supernova whose 
massive stellar progenitor had been been stripped of most, if not all, of its outermost 
hydrogen layer, but had retained its next-inner helium layer, before explosion. I will 
discuss the signiÞcance of this supernova, the derived size of its Wolf-Rayet progenitor, what it tells us about the explosive demise of massive stars, and its implications for 
the supernova-GRB connection. Furthermore, I will present observational results that 
confirm low metallicity as a key player in determining whether some massive stars die 
as GRB-SN or as an ordinary SN without a GRB. I show that the oxygen abundances at 
the SN-GRB sites are systematically lower than those found near ordinary broad-lined 
SN Ic, at a cut-off value of 0.3$-$0.5 Z$_{solar}$. 
\vfill\eject
\subsection{Rapid Spectrophotometric Changes in R127 and Reversal of the Decline (Nidia I. Morrell, Roberto C.  Gamen, Nolan R. Walborn, Rodolfo H. Barba, Katrien 
Uytterhoeven, Artemio Herrero, Christiopher Evans, Ian Howarth \& Nathan Smith)}
R127, the famous Luminous Blue Variable in the Large Magellanic Cloud,
was found in the peculiar early-B state and fainter in January 2008, suggesting
that the major outburst which started sometime between 1978 and 1980 was drawing
to a close, and that the star would presumably continue to fade and move to earlier spectral
types until reaching its quiescent Ofpe/WN9 state. Archival data showed that the main spectral 
transformation from the peculiar A-type state at maximum started between 2005 and 2007, and 
that it was in close concordance with features in the light curve.
However, subsequent observations during 2008 and early 2009 have shown
that the spectrum of R127 is now returning to a cooler, lower excitation
state, while the photometry shows a new brightening of the star.
A speculative 7-year cycle during the decline bears
further investigation. The curious behavior of R127 provides an opportunity to
gain further insight into the rapid transitional stages in the late
evolution of very massive stars.
\subsection{The Luminous Blue Variable Stars in M33: the Extended Hot Phase of Romano's Star (GR 290) (Corinne Rossi, Vito Francesco Polcaro, Silvia Galleti, Roberto Gualandi, Laura Norci \& Roberto F.  Viotti)} Romano's Star (GR290) is an LBV in M33. Recently, the Ê
star underwent a dramatic decrease in the visual, that was accompanied Ê
by a marked increase of the spectral line excitation. Presently, GR290 appears to be in the hottest phase ever observed in an LBV. More Ê
than 100 emission lines have been identified in the 3100$-$10000\AA\ range Ê
covered by the WHT spectra, including the hydrogen Balmer and Paschen Ê
series, He\.{\sc i} and He\,{\sc ii}, C\,{\sc iii}, N\,{\sc ii-iii}, Si\,{\sc iii-iv}, and many forbidden Ê
lines of [O\,{\sc iii}], [N\,{\sc ii}], [S\,{\sc iii}], [Ar\,{\sc iii}] and [Fe\,{\sc iii}]. Many lines, Ê
especially the He\,{\sc i} triplets, show a P Cygni profile with an E-A Ê
radial velocity difference of about 400 km/s. The 2008 spectrum appears Ê
quite similar to that of a typical WN8-9 star. During 2003$-$2009 GR290 Ê
varied between the WN11$-$WN8 spectral types, with the hottest spectrum Ê
corresponding to a fainter visual magnitude. This temperature-visual Ê
luminosity anticorrelation suggests variation at constant Mbol. GR290 Ê
might just present the key evidence that will help to bridge the LBV Ê
and WNL evolutionary phases.
\subsection{X-Ray modeling of Eta Car and WR140 from hydrodynamic simulations (Christopher
Russell, Michael F. Corcoran, Atsuo  Okazaki, Thomas I.Madura \& Stanley Owocki)}The colliding wind binary (CWB) systems Eta \ Car and WR140 provide unique laboratories for X-ray astrophysics. Their wind-wind collisions produce hard X-rays, which 
have been monitored extensively by several X-ray telescopes, such as RXTE and Chandra. To interpret these X-ray light curves and spectra, we apply 3D hydrodynamic 
simulations of the wind-wind collision using both smoothed particle hydrodynamics 
(SPH) and Þnite difference methods. We Þnd isothermal simulations that account for 
the absorption of X-rays from an assumed point source of X-ray emission at the apex 
of the wind-collision shock cone can closely match the RXTE light curves of both Eta \
Car and WR140. We are now applying simulations with self-consistent energy 
balance and extended X-ray emission to model the observed X-ray spectra. We present 
these results and discuss efforts to understand the earlier recovery of Eta \ Car's RXTE 
light curve from the 2009 minimum. 
\subsection{Accretion onto the secondary of Eta Car during the spectroscopic event (Noam Soker \& Amit Kashi)} We show that near periastron passage the shocked primary wind becomes gravitationally bound to the secondary star. This results in accretion flow onto the secondary star
that almost shuts down the secondary wind. The accretion process is the mechanism of 
the deep X-ray minimum. Not only in the present Eta \ Car, but also during the great 
eruption, accretion played a key role. 
\subsection{  New massive, eclipsing, double-lined spectroscopic binaries: Cyg OB2-17 \& NGC 346-13 (V. E. Stroud, J. S. Clark, I. Negueruela, D. J. Lennon \& C. J. Evans)}Massive, eclipsing, double-lined spectroscopic binaries are not common but necessary 
to understand the evolution of massive stars as they are the only direct way to determine 
the masses of OB stars and therefore obtain mass-luminosity functions. They are also 
the progenitors of energetic phenomena such as X-ray binaries and $\gamma$-ray bursts. 
We discuss results from photometric and spectroscopic studies of two binary systems: Cyg OB2-B17 which is a semidetached binary located in the Cyg OB2 association and comprised of 2 O supergiants; and NGC 346-13 which is a system located 
in the Small Magellanic Cloud and comprised of a semi-evolved B1 star and a hotter, 
optically fainter secondary, suggesting mass transfer in the system. 
\subsection{Monte Carlo radiative transfer in stellar wind
(Brankica Surlan \& Jiri Kubat)} 
As a first step towards solution of the radiative transfer equation in
clumped stellar wind we started to develop a code for the formal solution of 
the radiative transfer equation for given velocity, temperature, and density 
stratification. Wind structure was taken from a model calculated using a NLTE 
code by Krti\v{c}ka \& Kub\'at (2004, A\&A 417, 1003).
Wind opacity consists of line scattering under Sobolev approximation and
of the electron scattering. As our first preliminary results we plot the P~Cygni
profile of the line obtained from our calculation.
This work has been supported by grants 205/08/0003 and 205/08/H005 (GA \v{C}R).
\subsection{Gamma-ray observations of the Eta Car region (Marco Tavani, Sabina Sabatini, Roberto Viotti, Michael F.  Corcoran, Elena Pian \& the \textit{AGILE} Team} We present the results of extensive observations by the gamma-ray
\textit{AGILE} satellite of the Galactic region hosting the Carina nebula
and the colliding wind binary Eta \ Car. The \textit{AGILE} gamma-ray
satellite monitored the Eta \ Car region in several occasions
during the period 2007 July to  2009 January. \textit{AGILE} detects a gamma-ray
source consistent with the position of Eta \ Car. The average gamma-ray
flux above 100 MeV integrated over the pre-periastron period
2007 July - 2008 October is F = (37 +/- 5)$\times$10$^{-8}$ ph/cm$^2$/sec
corresponding to an average gamma-ray luminosity of L = 3.4$\times$10$^{34}$ erg/sec
for a distance of 2.3 kpc. \textit{AGILE} also detected a remarkable 2-day
gamma-ray flaring episode of on 11-13 October 2008, most likely caused by
a colliding wind transient particle acceleration episode. The pre-periastron
gamma-ray emission appears to be erratic, and is possibly related to
transient acceleration and radiation episodes in  the strongly  
variable colliding wind shocks in the system.
Our results provide the long sought first detection above 100 MeV
of a colliding wind binary, and have important theoretical
implications.
 \vfill\eject\subsection{Long-term variability of Eta Car (Mairan Teodoro)} During the last 50 years, Eta \ Car has increased its brigthness at variable rates. 
For instance, the central source presented V=8 from 1910 to 1940, when it suddenly 
increased its brightness by 1 magnitude in a few years. Since then, the brightness 
has increased almost linearly with time at a rate of approximately 0.03 mag per year. 
However, after the spectroscopic event of 1997.9, the rate increased to 0.2 mag per year 
and remained so until mid-2006, when a drop in the brightness of the central source 
was observed (almost 30 per cent in less than one year!). In this work we present 
the results of our study on the long-term variability of the central source of Eta Car, 
showing that, while the central source is getting brighter, the equivalent width of the 
lines are getting weaker from cycle to cycle. Besides, our results indicate that at least 
in the last 4 events, the behaviour of the high- and intermediary-excitation lines near 
the spectroscopic event have not changed signiÞcantly. 
\subsection{Eta Car around the 2009 periastron - a new view with X-shooter (Christina Thone, Theodore R.  Gull, Guido Chincarini, Elena Pian, Henrik Hartman, Sandro D'Odorico \& Lex Kapor)}
We observed the Eta \ Car binary system with the X-shooter 
spectrograph at the VLT during commissioning phase that spanned the latest  periastron event of the system on 
Jan. 11 2009. X-shooter covers the whole spectral range from the UV (3000\AA\ to the 
IR (2.5 $\mu$m) simultaneously with medium resolution ($R=\lambda/\delta\lambda=4000-9000$). Two long slits were 
placed on the Homunculus skirt  radially extending out from the star in opposite 
directions at three different epochs in January (5Ð10 d after periastron), March and June. 
At visible wavelengths, the Strontium Filament was sampled with three sub-slits of the 1.8"
$\times$4" Integral Field Unit (IFU) in January. 
The shape of the Balmer lines in the opposite slit positions can give us information about 
the orientation of the orbit of the secondary star. The absence of PCygni absorption 
on the south-west slit indicates that the secondary enters from the south-western side 
ionizing the wind material causing the absorption in the north-east slit. The X-ray 
emission, which disappears during periastron due to the collapse of the shock front of 
the winds, recovered surpisingly early in 2009. High ionization lines were still not visible 
again in the data of the March run while they are still visible in the outer regions of the 
radial slits in January since those regions had not yet seen the shut off of the FUV radiation due 
to the light travel time. 
\subsection{INTEGRAL observations of Eta Car (Roland Walter \& Jean-Christophe Leyder)} If relativistic particle acceleration takes place in colliding-wind binaries, then hard X-rays and $\gamma$-rays are expected 
through inverse Compton scattering of the copious UV radiation field.
The \textit{INTEGRAL} satellite provided hard X-ray images of the Carina region with a much higher spatial resolution
than previously available. Based on observations taken far from periastron, a bright source was detected at the position 
of  Eta ~Car  up to 100~keV. Two additional nearby hard X-ray sources could also be resolved. This is the first unambiguous 
detection of Eta ~Car at hard X-rays. There is no other X-ray source in the hard X-ray error circle, bright enough to match 
the hard X-ray flux.\\
\indent The average hard X-ray emission of Eta  Car in the 22-100~keV energy range is very hard (with a photon index $\Gamma\approx 1$) 
and its luminosity ($7\times 10^{33}$erg/s) is in agreement with the predictions of inverse Compton models and corresponds to
about 0.1\% of the energy available in the wind collision. \\\indent New \textit{INTEGRAL} observations were taken during the 2009 periastron passage, and the first results are presented. 
Only a 5-$\sigma$ upper-limit could be derived. This is consistent with a lower fraction of very energetic particles during 
periastron than outside. This could perhaps be linked with electron cooling by the extreme radiation field.\subsection{ \textit{BRITE-Constellation} (Werner W. Weiss, Anthony F. Moffat \& the \textit{BRITE-Constellation} Team)}\textit{BRITE-Constellation}, a project developed since 2003 by researchers at Canadian and 
Austrian Universities presently consists of \textit{UniBRITE} and \textit{BRITE-Austria/TUG-SAT1}, 
which are two 20 cm cube nanosatellites. Each will host a 30 mm aperture telescope 
with a CCD camera equipped with either a red (550 to 700 nm) or a blue (390 to 
460 nm) Þlter, to perform high-precision two-color photometry of the brightest stars 
in the sky for up to several years. Depending on the orbit and the position of the 
\textit{BRITE} targets the photometry can be obtained contiguously during many orbits for 
many months, with gaps during individual orbits, or only for certain periods of the 
year. \\
\indent The primary science goals are studies of luminous stars in our neighbourhood, representing objects which dominate the ecology of our Universe, and of evolved stars to 
probe the future development of our Sun.\\
\indent A launch of  \textit{UniBRITE} and \textit {BRITE-Austri} in 2009 is envisioned and an expansion proposal of the \textit {BRITE-Constellation} by two additional spacecraft of the same 
construction is currently under review in Canada.

\end{document}